\documentclass[a4paper,fleqn,usenatbib]{mnras}
\usepackage{graphics}
\usepackage{graphicx}
\usepackage{psfig}
\input{epsf}
\usepackage{amsmath}
\usepackage{amssymb}
\usepackage{url}
\usepackage[T1]{fontenc}
\usepackage{ae,aecompl}
\usepackage{widetext}

\newcommand{\order}[1]{\mathcal{O}\!\left(#1\right)}

\newcommand{\ud}{\mathrm{d}}

\title[Clustering of Local Extrema in {\it Planck} maps] {Clustering of Local Extrema in {\it Planck} CMB maps}

\author[A. Vafaei Sadr \& S. M. S. Movahed]{
A. Vafaei Sadr$^{1}$ \& S. M. S. Movahed$^{2,1}$\thanks{E-mail: m.s.movahed@ipm.ir}
\\
$^{1}$ School of Physics, Institute for Research in Fundamental Sciences (IPM), P. O. Box 19395-5531, Tehran, Iran\\
$^{2}$Department of Physics, Shahid Beheshti University,  1983969411, Tehran, Iran
}
\begin{document}
\maketitle

\begin{abstract}

The clustering of local extrema will be exploited to examine Gaussianity, asymmetry, and the footprint of the cosmic-string network on the CMB observed by {\it Planck}. The number density of local extrema ($n_{\rm  pk}$ for peak and $n_{\rm tr}$ for trough) and sharp clipping ($n_{\rm   pix}$) statistics support the Gaussianity hypothesis for all component separations. However, the pixel at the threshold reveals a more consistent treatment with respect to end-to-end simulations. A very tiny deviation from associated simulations in the context of trough density, in the threshold range $\theta\in [-2-0]$ for \texttt{NILC} and \texttt{CR} component separations, are detected. The unweighted two-point correlation function,  of the local extrema illustrates good consistency between different component separations and corresponding Gaussian simulations for almost all available thresholds. However, for high thresholds, a small deficit in the clustering of peaks is observed with respect to the {\it Planck} fiducial $\Lambda$CDM model. To put a significant constraint on the amplitude of the mass function based on the value of $\Psi$ around the Doppler peak ($\theta\approx 70-75$ arcmin), we should consider $\vartheta\lesssim 0.0$.  The scale-independent bias factors for the peak above a threshold for large separation angle and high threshold level are in agreement with the value expected for a pure Gaussian CMB. Applying the $n_{\rm pk}$, $n_{\rm tr}$, $\Psi_{\rm pk-pk}$ and $\Psi_{\rm tr-tr}$ measures on the tessellated CMB map with patches of  $7.5^2$ deg$^2$ size prove statistical isotropy in the {\it Planck} maps. The peak clustering analysis puts the upper bound on the cosmic-string tension, $G\mu^{(\rm up)} \lesssim 5.59\times 10^{-7}$, in \texttt{SMICA}.
\end{abstract}

\begin{keywords}
methods: data analysis - methods: numerical - methods: statistical - cosmic microwave background - theory - early Universe.

\end{keywords}

\section{Introduction}
Cosmological stochastic fields are ubiquitous in various
observations. Any conceivable theory incorporating initial
conditions \citep{Malik:2008im}, cosmic microwave background (CMB)
\citep{dodelson2003modern,lesg13,Lesgourgues:2013qba,Ade:2015xua},
large scale structures \citep{Bernardeau:2001qr,Cooray:2002dia} and
other relevant fields \citep{Kashlinsky:2004jt,Lewis:2006fu}
essentially includes stochastic notion. The initial conditions
and/or evolution for cosmological fields are specified with random
behaviour. To infer any reliable bridge between model building and
observational quantities, it is necessary to use robust statistical
tools. By means of the central limit theorem and statistical
isotropy, it is possible to use a perturbative approach to
characterize the stochastic field
\citep{Matsubara:2003yt,Codis:2013exa,matsubara2020statistics}.

Many topological and geometrical measures have been introduced to characterize morphology of cosmological stochastic fields, $\mathcal{F}$, in $1+1$, $1+2$ and $1+3$ dimensions\footnote{According to the {\it measure theoretic} approach which is ultimately identical to the  probabilistic description, a typical $(n+D)$-Dimensional stochastic field, $\mathcal{F}^{(n,D)}$, is a measurable mapping from probability space into a $\sigma$-algebra of $\mathbb{R}^{n}$-valued function on $\mathbb{R}^D$ Euclidian space \citep{adler81,adler2011topological,adler2010persistent}. Here,  the index $n$ refers to $n$-dependent parameters and $D$ represents $D$-independent parameters describing a $(n+D)$-Dimensional random field or a stochastic process.}. Critical and excursion sets are generally the backbone for the definition of more significant features on a smoothed stochastic field. Critical sets include features incorporating conditions for having local and extended extrema \citep{Matsubara:2003yt,Pogosyan:2008jb,Gay:2011wz,Codis:2013exa,matsubara2020statistics}. A rigorous definition for excursion sets for a function of given stochastic field, $\mathcal{F}(X)$, above a threshold $\vartheta$ is defined by: $\mathcal{A}_{\vartheta}(\mathcal{F})\equiv\{X|\mathcal{F}(X)\ge \vartheta \}$ \citep{adler81}. Accordingly, in real or harmonic space, we are able to achieve theoretical descriptions of the corresponding features in a cosmological stochastic field irrespective of its dimension.  The mentioned benchmarks have advantages and disadvantages from theoretical and computational points of view. Complicated algorithms and marginal behaviour  with respect to an arbitrary exotic feature are some of the disadvantages. Nevertheless, there are many benefits to setting up such estimators beyond standard methods. Among them are the magnification of deviation and  capability of  discriminating the exotic features embedded in a typical cosmological field.

One-point statistics provides considerable information regarding the abundance of the underlying features while the complex nature of cosmological stochastic fields  is essentially going beyond a one-point analysis. To characterize such complexity, we should take into account much more complicated behaviour in precise observations. Subsequently, $N$-point correlation functions of  arbitrary features are therefore common estimators. In the context of Two-Point Correlation Function (TPCF), there are two relevant measures to asses clustering:
I) the weighted TPCF deals with  the autocorrelation and II)
the unweighted TPCF estimates  the excess probability of finding the  pair of features by imposing proper conditions for a given separation distance (or e.g. time, angle) \citep{peeb80,kaiser1984spatial,peac85,lumusden89,Bardeen:1985tr,Bond:1987ub,davis_peeb83,hamilton1993toward,szalay88,hewet82,Landy:1993yu,Marcos-Caballero:2015lxp}. In principle, there is a systematic relation between both TPCFs \citep{rice1954selected,taqqu1977law,kaiser1984spatial,szalay88,desjacques2010modeling,desjacques2018large}.

Level crossing statistics is a pioneering approach for characterizing stochastic processes introduced by S. O. Rice \citep{rice44a,rice44b}. Up-, down- and conditional crossing statistics are modifications to primary definition of level crossing \citep{Bardeen:1985tr,Bond:1987ub,ryden1988,ryd89,mat96a,percy00,Matsubara:2003yt,tabar03,sadegh11,sadegh15}.  Minkowski functionals which are also closely related to the crossing statistics provide $1+D$ functionals to quantify morphology in $D$ dimension \citep{hadwiger2013vorlesungen}  and have been utilized for cosmological random fields \citep{mecke1994robust,schmalzing1995minkowski,schmalzing1998minkowski,Matsubara:2003yt,matsubara2010analytic,Hikage:2006fe,Codis:2013exa,ling2015distinguishing,fang2017new}. A number of critical sets including peaks (hills), troughs (lakes), saddles, voids, skeleton, genus and Euler characteristics, are more popular in cosmology for different purposes and  they have been fully explored for Gaussian stochastic fields. Some extensions for non-Gaussian and anisotropic conditions have been done in some research \citep{Matsubara:2003yt,Pogosyan:2008jb,Pogosyan:2011qq,Gay:2011wz,Codis:2013exa}.  More recently, Betti numbers, Euler characteristic and Minkowski functionals for a set of cosmological 3D fields have been examined extensively \citep{pranav2019topology}. The scaling approach for investigating cosmological stochastic fields has been discussed by \cite{Borgani:1994uy,SadeghMovahed:2006em}. Standard estimators like three- and four-point functions in real space, bispectrum and trispectrum in harmonic space, multiscaling methods such as wavelet \citep[and references therein]{2016A&A...594A..17P,ade2014plancknon} and regenerating of stochastic process based on the  Fokker-Planck equation \citep{ghasemi2006characteristic} have been also considered.

The CMB has a stochastic nature encoded by various phenomena ranging from high-energy and primordial
events to low-energy scales \citep{dodelson2003modern,lesg13,Lesgourgues:2013qba,Ade:2015xua}.  Some relevant topics for examining the CMB by statistical tools are as follows:  different anomalies \citep{planck2013results,Ade:2015hxq}, non-Gaussianity \citep{ade2014plancknon,2016A&A...594A..17P,Renaux-Petel:2015bja,Ade:2015hxq,larson2004hot,Larson:2005vb,2009MNRAS.396.1273H} and other exotic phenomena \citep{sadegh11,Movahed:2012zt,Ade:2013xla,sadr2018cosmic,vafaei2017multiscale}.

Peaks and pixels statistics are proper measures of CMB in one- and two-point forms and have been extensively used to investigate CMB data released by various surveys  \citep{sazhin1985hot,Bond:1987ub,nicola87,Cayon:1995ms,Fabbri:1995md,kogut95,kogut96,Barreiro:1996ds,Heavens:1999cq,Heavens:2000mu,Kashlinsky:2001zla,Futamase:2000qb,Dore:2002xm,HernandezMonteagudo:2002df,2009MNRAS.396.1273H,Tojeiro:2005mt,larson2004hot,Larson:2005vb,Rossi:2009wm,Rossi:2010hu,Pogosyan:2011qq,Movahed:2012zt,Rossi:2013fea,Ade:2015hxq,sadr2018cosmic,vafaei2017multiscale}. Two-Dimensional  topology, which is related to the statistics of the hot- and coldspots of the underlying field has been evaluated for CMB random field \citep{colley2003genus,Gott:2006za,Colley:2014nna,Ade:2015hxq}. Persistent homology in the context of Topological Data Analysis has been considered for searching the non-Gaussianity in the CMB map \citep{cole2018persistent}. In addition, the detectability of gravitational lensing in the CMB map based on the TPCF of hotspots has been examined by \cite{takada2000gravitational,takada2001detectability}.

Searching in various works reveal that the clustering evaluation in CMB maps for different purposes has mostly been  concentrated on statistics of regions above or below a threshold without taking into account proper conditions on the first- and second-derivatives properties of the underlying field \citep{Kashlinsky:2001zla,Rossi:2009wm,Rossi:2010hu,Rossi:2013fea}. The best-fitting values of the spectral parameters that  are necessary to determine the theoretical prediction of probability distribution of peaks for a Gaussian random field have been computed \citep{Ade:2015hxq}. Also, in the context of one point statistics, their results confirmed the consistency with Gaussian property. In spite of extensive analysis of peaks distributions and location by {\it Planck} team, going beyond one-point statistics, provides more useful information on the nature of clustering and probing exotic features.

 In this paper, however we will focus on the clustering of local extrema to examine whether such critical sets are more sensitive to declaring  exotic features embedded in the CMB map as well as their robustness in the presence of noise.
 Here, we deal with the local extrema statistics of the {\it Planck} CMB maps to study the following main objects and novelties:\\
1) The consistency between various component separation pipelines leading to different observed map will be checked by considering the clustering of local extrema. We will carry out some robust estimators to compute local extrema clustering, as they are supposed to be free of the boundary effect. \\
 2) We will also probe the non-Gaussianity based on the capability of local extrema clustering. Meanwhile, we will also modify the series expansion for the number density of sharp clipping. \\
 3) The scale-dependent and independent  bias factors according to the general definition of bias will be determined for different components.\\
4) The asymmetry of the CMB according to the unweighted TPCF of peaks, troughs, and number density of local extrema  for various component separations will be examined.\\
5) We will compute the value of the upper bound on the cosmic string tension, by comparing clustering of critical sets computed for pure Gaussian CMB maps including all foreground residuals, systematic noises and beam effects and those induced by a cosmic string network simulated according to numerical simulations of Nambu-Goto string networks using the Bennett-Bouchet-Ringeval code~\citep{Bennett:1990,Ringeval:2005kr}.

The rest of this paper is organized as follows: in Section \ref{stat}, mathematical description of local extrema statistics will be clarified. The statistical definition of bias factor will be given in this section. Data description will be given in Section \ref{sec:data}. We will implement geometrical and topological measures on the synthetic and real CMB data in Section \ref{simul}. The Gaussian and asymmetry hypothesis and searching the cosmic-strings network on the CMB map in  Section \ref{simul}. The last section will be devoted to summary and conclusions.

\section{Theoretical Notions}\label{stat}
The statistics of local extrema (both minima and maxima) provides a robust framework to search for evidence of non-Gaussianity in  data \citep{Matsubara:2003yt,Tojeiro:2005mt,Pogosyan:2008jb,Pogosyan:2011qq,matsubara2020statistics} and to look for exotic features such as topological defects (e.g. cosmic strings network) \citep{Heavens:1999cq,Heavens:2000mu,Movahed:2012zt,vafaei2017multiscale,sadr2018cosmic}.  Such an extremum is defined as a pixel whose amplitude is either higher or lower than the adjacent nearest neighbours incorporating conditions on the first and second derivatives of field. Therefore, we have additional mathematical conditions when we deal with extrema compared to sharp clipping.

For a statistically isotropic Gaussian stochastic field, the number density of peaks was derived by \cite{Bond:1987ub}. The non-Gaussian extrema counts for the CMB field have been studied by \cite{Pogosyan:2011qq}. It has been expressed that according to the perturbation approach for a smooth non-Gaussian field, it is possible to track different shapes of non-Gaussianity \citep{Pogosyan:2011qq}. However, \cite{Movahed:2012zt} showed that the footprint of non-Gaussianity produced by cosmic-strings network cannot be recognized by utilizing only number counts of extrema. Subsequently, we conclude that clustering of coldspots and hotspots manifested by extrema outliers in the trough and peak values can constitute evidence for non-Gaussianity or deviation from isotropy \citep{Ade:2015hxq,akrami2020planck}. \\

\begin{figure}
\begin{center}
\includegraphics[width=0.7\columnwidth]{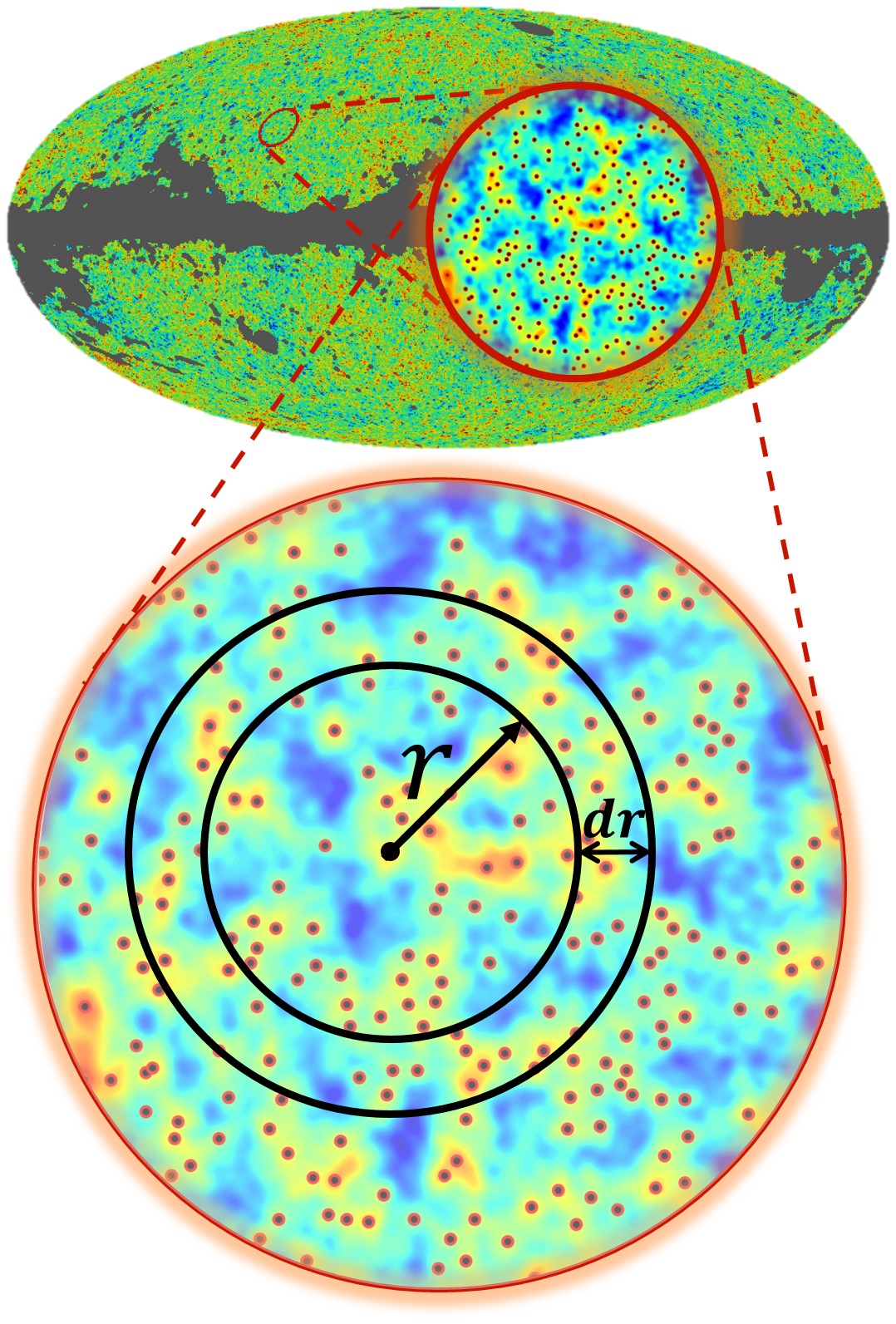}
\caption{Peaks distribution on the \texttt{NILC} map for $N_{\rm side}=512$ at threshold $\vartheta=0.5$. In the enlarged plot, we indicate a sketch to illustrate clustering of local peaks separated by $r$.}
\label{fig:onlypeaks}
\end{center}
\end{figure}

\subsection{Local extrema counts and excursion sets}
 For the sake of clearance, we will specify number density of local extrema, number density of regions above (below) a threshold and unweighted two-point correlation function of critical sets in the probabilistic framework for the CMB map. Both observed and simulated CMB temperature anisotropy maps are stochastic fields represented  by  the 2-dimensional maps, $T\in{\rm L}^2(\mathbb{R}^2)$. We define the vector $\mathcal{A}$ at each spatial point denoted  by $(\theta,\phi)$ on the CMB map by:
$$\mathcal{A}_{\mu}\equiv\{\delta_T,\eta_{\phi},\eta_{\theta},\xi_{\phi\phi},\xi_{\theta\theta},\xi_{\phi\theta}\}$$
where $\delta_T\equiv\Delta T(\theta,\phi)/T(\theta,\phi)$ is the temperature fluctuation, $\eta_{\phi}\equiv \partial \delta_T/\partial \phi$, $\eta_{\theta}\equiv \partial \delta_T/\partial \theta$ and $\xi_{\phi\theta}\equiv\partial^2 \delta_T/\partial \phi\partial \theta$. To examine local extrema, we therefore need the first- and second-order derivatives. To determine the joint probability density function (PDF) of $\mathcal{A}$, which is essential to compute the average of any features,  we use the so-called characteristic function  defined by:
\begin{eqnarray}
\mathcal{Z}_{\mathcal{A}}(\lambda)=\int_{-\infty}^{+\infty}\ud^6{\mathcal{A}}
  {\mathcal{P}}({\bf{\mathcal{A}}}) \bf{e}^{{\bf{i}}\lambda.{\bf{\mathcal{A}}}}
\end{eqnarray}
where $\lambda$ is an array with the same size as $\mathcal{A}$. The perturbative expansion of $\mathcal{Z}_{\mathcal{A}}$ becomes \citep{Matsubara:2003yt}:
\begin{eqnarray}\label{parti1}
&&\mathcal{Z}_{\mathcal{A}}(\lambda)=\exp{ \left( {-\frac{1}{2}\lambda^T.\mathcal{K}^{(2)}.\lambda}\right)}\nonumber\\
 &&\times \exp{\left[ \sum_{j=3}^{\infty}\frac{{\bf{i}}^j}{j!}\left(\sum_{\mu_1}^N\sum_{\mu_2}^N...\sum_{\mu_j}^N \mathcal{K}^{(j)}_{\mu_1,\mu_2,...,\mu_j}\lambda_{\mu_1}\lambda_{\mu_2}...\lambda_{\mu_j}\right) \right] } \nonumber\\
\end{eqnarray}
where ${\mathcal K}^{(n)}_{\mu_1,\mu_2,...,\mu_n}\equiv\langle \mathcal{A}_{\mu_1}\mathcal{A}_{\mu_2}...\mathcal{A}_{\mu_n} \rangle$  is the  array of connected cumulants. Also $\mathcal{K}^{(2)}\equiv \langle {\mathcal A} \otimes{\mathcal A}\rangle$ represents the $6\times6$ covariance matrix of ${\mathcal A}$ at each spatial point. In the appendix, we give the details for the form of covariance matrix elements.  The joint probability density function of the CMB map including the
higher order derivatives of $\delta_{T}$ can be inferred by the
inverse Fourier transform of the characteristic function (Eq.
(\ref{parti1})) as:
\begin{equation}\label{pdf1}
\begin{aligned}
  \mathcal{P}({\mathcal{A}})=
  \exp \bigg[\sum_{j=3}^{\infty}\frac{(-1)^j}{j!}
      \bigg(\sum_{\mu_1=1}^6 &...\sum_{\mu_j=1}^6 {\mathcal
        K}^{(j)}_{\mu_1,\mu_2,...,\mu_j} \\ & \times \frac{\partial^j}{\partial \mathcal{A}_{\mu_1}...\partial \mathcal{A}_{\mu_j}}\bigg)\bigg] {\mathcal {P}}_{\rm G}({\mathcal{A}})\
\end{aligned}
\end{equation}
where $ {\mathcal {P}}_{\rm G}({\mathcal{A}})=\frac{1}{\sqrt{(2\pi)^{6} |\mathcal{K}^{(2)}|}} \
{\bf{e}}^{-\frac{1}{2}({\bf \mathcal{A}}^{T}.[\mathcal{K}^{(2)}]^{-1}.{\bf
    \mathcal{A}})}$. The perturbative form of the one-point PDF of the temperature fluctuations,
 $\mathcal{P}_{\delta_T}(\alpha)$, in the presence of weak non-Gaussianity has been derived in \cite{vafaei2017multiscale}.
Subsequently, theoretical definition of local extrema number density at a given threshold, $\delta_T\equiv\alpha=\vartheta\sigma_0$, is \citep{Bardeen:1985tr}:
\begin{eqnarray}\label{number1}
n_{\diamond}(\vartheta)&\equiv&\langle n_{\diamond}(\vartheta;{\bf r})\rangle=\langle \delta_{\rm D}({\bf{r}}-{\bf{r}}_{\diamond})\rangle\nonumber\\
&=& \int {\rm d}^6{\mathcal{A}}  \delta_{\rm D}({\bf{r}}-{\bf{r}}_{\diamond}) {\mathcal{P}}({\mathcal{A}}_{\mu})
\end{eqnarray}
where $\diamond$ can be replaced by "$\rm pk$" for peak, "$\rm tr$" for trough and "$\rm pix$" for sharp clipping. The ${\bf{r}}_{\diamond}$ represents the location vector of local extrema and sharp clipping on the CMB map at threshold $\vartheta$.  $\delta_{\rm D}$ is the Dirac delta function. Clarifying the relation between Dirac delta function and ${\rm d}^6{\mathcal{A}}$ enables us to write the average of local extrema number density at the threshold as:
 \begin{eqnarray}
n_{\diamond}(\vartheta)=\langle \delta_{\rm D}(\delta_T-\vartheta\sigma_0)\delta_{\rm D}(\eta_{\phi})\delta_{\rm D}(\eta_{\theta})|{\rm det}(\xi)|\rangle
\end{eqnarray}
The second-derivative tensor of the CMB field ($\xi_{ij}$) should be
{\it negative definite} ({\it positive definite}) at peak (trough)
position. Finally, the number density of peaks for a purely
isotropic Gaussian CMB field in the threshold interval,
$[\vartheta,\vartheta+d\vartheta]$, becomes
\citep{Bardeen:1985tr,Bond:1987ub}:
\begin{equation}\label{nu1}
n_{\rm pk}({\vartheta})=\frac{N_{\rm pix}^{\rm tot}}{4\pi}\frac{e^{-{\vartheta}^2/2}{\mathcal{G}}({\Gamma},{\Gamma}{\vartheta})}{(2\pi)^{3/2}\gamma^2}
\end{equation}
where
\begin{eqnarray}\label{g1}
&&{\mathcal{G}}({\Gamma},{\Gamma}{\vartheta})\equiv ({\Gamma}^2{\vartheta}^2-\Gamma^2)\left\{1-\frac{1}{2}{\rm erfc} \left[ \frac{{\Gamma}{\vartheta}}{\sqrt{2(1-\Gamma^2)}}\right]\right\}\nonumber\\
&&+{\Gamma}{\vartheta}(1-\Gamma^2)\frac{e^{-\frac{{\Gamma}^2{\vartheta}^2}{2(1-\Gamma^2)}}}{\sqrt{2\pi(1-\Gamma^2)}}\nonumber\\
&&+\frac{e^{-\frac{{\Gamma}^2{\vartheta}^2}{3-2\Gamma^2}}}{\sqrt{3-2\Gamma^2}}\left\{ 1-\frac{1}{2}{\rm erfc}\left[\frac{{\Gamma}{\vartheta}}{\sqrt{2(1-\Gamma^2)(3-2\Gamma^2)}}\right]\right\}\nonumber\\
\end{eqnarray}
in which $\rm erfc(:)$ stands for the complementary error function. The parameters $\Gamma$ and $\gamma$ in Eqs. (\ref{nu1}) and (\ref{g1}) are defined by:
$\Gamma\equiv \frac{\sigma_1^2}{\sigma_0\sigma_2}$ and $\gamma\equiv\sqrt{2}\frac{\sigma_1}{\sigma_2}$ \citep{Bond:1987ub}. $\Gamma\in[0,1]$ characterized  the shape of power spectrum, while $\gamma$ indicates the characteristic radius of local extrema. Also $N_{\rm pix}^{\rm tot}=12N^2_{\rm side}$ represents the total number of pixel in a given map with resolution specified by $N_{\rm side}$ computed by \texttt{HEALPIX} software \citep{Gorski:2004by}. The various orders of spectral indices are given by:

\begin{eqnarray}\label{eq:spectral}
\sigma_m^2= \sum_{\ell}\frac{(2\ell+1)}{4\pi}\left[\ell(\ell+1)\right]^m C_{\ell}^{TT} W^2_{\ell}
\end{eqnarray}
where $W_{\ell}$ is beam function and $C_{\ell}^{TT}$ is temperature power spectrum.
For the sharp clipping above (below) a threshold corresponding to the pixels above (below) a threshold, we do not take into account  the constraints on the first- and second-derivative of the underlying field and therefore, we have $n_{\rm pix}(\vartheta)=\langle \Theta(\delta_T\mp\vartheta\sigma_0)\rangle$ ($\Theta$ is step function). The minus (plus) sign is for above (below) threshold. According to  Eq. (\ref{pdf1}), the perturbative number density of pixels above a threshold in the non-Gaussian field (NG) reads as:
\begin{eqnarray}\label{eq:pixnon}
&&n_{\rm pix}^{\rm NG}(\alpha>\vartheta\sigma_0)\equiv \langle \Theta(\delta_T-\vartheta \sigma_0 )\rangle= \frac{N_{\rm pix}^{\rm tot}}{4\pi}\frac{1}{2} \text{erfc}\left(\frac{\vartheta }{\sqrt{2}}\right)\nonumber\\
&&+\frac{N_{\rm pix}^{\rm tot}}{4\pi}\left[\frac{e^{-\frac{\vartheta ^2}{2}}\left(\vartheta ^2-1\right) S_0}{6 \sqrt{2 \pi }}\right]\sigma_0\nonumber\\
&&+\frac{N_{\rm pix}^{\rm tot}e^{-\frac{\vartheta ^2}{2}}}{4\pi}\left[\frac{3 K_0\vartheta(\vartheta^2-3)+S_0^2\vartheta(\vartheta ^4-10 \vartheta^2+15)}{72 \sqrt{2 \pi }}\right]\sigma_0^2 \nonumber\\
&&+\mathcal{O}(\sigma_0^3)
\end{eqnarray}
where $S_0\equiv\mathcal{K}^{(3)}_{111}/\sigma_0^4$ and $K_0\equiv\mathcal{K}^{(4)}_{1111}/\sigma_0^6$. Eq. (\ref{eq:pixnon}) represents  a generalized form compared to  the one given by \cite{Rossi:2010hu}. Taking into account sharp clipping statistics up to $\mathcal{O}(\sigma_0)$ results in marginal behaviour for non-Gaussianity at $\vartheta=\pm1$. Including the  various spectral indices, $\sigma_m$, yields more complicated theoretical formula for $n_{\rm pk}$ and $n_{\rm tr}$, and mentioned measures are more sensitive to non-Gaussianity.  Perturbative expansion in the weakly non-Gaussian regime for number density of peaks and troughs have been calculated in \citep{Pogosyan:2011qq}. In Fig.  \ref{fig:onlypeaks}, we show the spatial distribution of peaks on the \texttt{NILC} map. In the next subsection, we will try to set up the clustering of local extrema which is systematically given by unweighted TPCF of peaks and troughs.

\subsection{Unweighted Two-Point Correlation Function}\label{TPCFtheory}
The one-point statistics of some geometrical measures (number density of local extrema as well as pixels above a threshold) have been explained in previous subsection. They can explore  probable exotic features and various types of non-Gaussianity \citep{Pogosyan:2011qq,Rossi:2010hu,Rossi:2013fea,Gay:2011wz,Codis:2013exa,Reischke:2015jga}. However, to do more precise evaluation,  we should go beyond the one-point statistics \citep{2009MNRAS.396.1273H,Movahed:2012zt}. In this subsection, we focus on the clustering of local extrema, which is the so-called unweighted TPCF.  Semi-analytical \citep{Heavens:1999cq,Heavens:2000mu,matsubara2020statistics} and numerical approaches  \citep{kerscher2000comparison} are often used to study the clustering of local extrema. The clustering of peak or trough pairs separated by distance $r=|{\bf{r}}_1-{\bf{r}}_2|$ at thresholds $\vartheta_1$ and $\vartheta_2$ is given by:
\begin{eqnarray}\label{eq:tpcf_peak}
&&\langle n_{\diamond}(\textbf{r}_1,\vartheta_1) n_{\diamond}(\textbf{r}_2,\vartheta_2) \rangle=\\
&& \int {\rm d}^6{\mathcal A}_{1} {\rm d}^6{\mathcal A}_{2}\delta_{\rm D}(\textbf{r}_1-\textbf{r}_{\diamond1})\delta_{\rm D}(\textbf{r}_2-\textbf{r}_{\diamond2}){\mathcal P}({\mathcal A}_{1};{\mathcal A}_{2})\nonumber
\end{eqnarray}
Taking into account the conditions for having extrema imposes some constraints on the domain of integrations \citep{Heavens:1999cq,Heavens:2000mu}. The excess probability of finding pairs using Eq. (\ref{eq:tpcf_peak}) becomes:
\begin{equation}
\Psi_{ \diamond-\diamond}(r;\vartheta_1,\vartheta_2)=\frac{\left \langle n_{\diamond}(\textbf{r}_1,\vartheta_1) n_{\diamond}(\textbf{r}_2,\vartheta_2)
  \right \rangle}{n_{\diamond}(\vartheta_1)n_{\diamond}(\vartheta_2)} -1,
\end{equation}
In our pipeline, we rely on the numerical evaluation of unweighted TPCF  of local extrema in both observed and simulated maps. To this end, some robust numerical estimators for determining $\Psi_{\diamond-\diamond}(r;\vartheta_1,\vartheta_2)$ that are free of the boundary effect are listed below:
\begin{widetext}
\begin{eqnarray} \label{eq:pp-estimator1}
\Psi^N_{\diamond-\diamond}(r;\vartheta_1,\vartheta_2)= \left(\frac{D_{\diamond}(\textbf{r}_1,\vartheta_1)D_{\diamond}(\textbf{r}_2,\vartheta_2)}{R_{\diamond}(\textbf{r}_1,\vartheta_1)R_{\diamond}(\textbf{r}_2,\vartheta_2)} \right) \frac{N_R^{\diamond}(N_R^{\diamond}-1)}{N_D^{\diamond} (N_D^{\diamond} -1)} -1\nonumber\\
\end{eqnarray}
\begin{eqnarray} \label{eq:pp-estimator2}
\Psi^H_{\diamond-\diamond}(r;\vartheta_1,\vartheta_2)= \frac{R_{\diamond}(\textbf{r}_1,\vartheta_1)R_{\diamond}(\textbf{r}_2,\vartheta_2)D_{\diamond}(\textbf{r}_1,\vartheta_1)D_{\diamond}(\textbf{r}_2,\vartheta_2)}{\left[D_{\diamond}(\textbf{r}_1,\vartheta_1)R_{\diamond}(\textbf{r}_2,\vartheta_2)\right]^2} -1\nonumber\\
\end{eqnarray}
\begin{eqnarray} \label{eq:pp-estimator3}
&&\Psi^{LS}_{\diamond-\diamond}(r;\vartheta_1,\vartheta_2)=\left( \frac{D_{\diamond}(\textbf{r}_1,\vartheta_1)D_{\diamond}(\textbf{r}_2,\vartheta_2)}{R_{\diamond}(\textbf{r}_1,\vartheta_1)R_{\diamond}(\textbf{r}_2,\vartheta_2)} \right) \frac{N_R^{\diamond} (N_R^{\diamond} -1)}{N_D^{\diamond} (N_D^{\diamond} -1)}  - \left( \frac{D_{\diamond}(\textbf{r}_1,\vartheta_1)R_{\diamond}(\textbf{r}_2,\vartheta_2)}{R_{\diamond}(\textbf{r}_1,\vartheta_1)R_{\diamond}(\textbf{r}_2,\vartheta_2)} \right) \frac{N_R^{\diamond} (N_R^{\diamond} -1)}{N_D^{\diamond} N_R^{\diamond}} +1\nonumber\\
\end{eqnarray}
\end{widetext}
$\Psi^N$ is called the 'natural estimator' \citep{Landy:1993yu}.  $\Psi^H$ is proposed by  \cite{hamilton1993toward} while $\Psi^{LS}$ introduced by  \cite{Landy:1993yu} has a nearly Poisson variance. In the above equation, $D_{\diamond}(\textbf{r}_1,\vartheta_1)D_{\diamond}(\textbf{r}_2,\vartheta_2)$ and $R_{\diamond}(\textbf{r}_1,\vartheta_1)R_{\diamond}(\textbf{r}_2,\vartheta_2)$ represent  the number of peak or trough pairs in the data  and random catalogs, respectively, and  $D_{\diamond}(\textbf{r}_1,\vartheta_1)R_{\diamond}(\textbf{r}_2,\vartheta_2)$ corresponds to cross-pairs. In the above equations, $N_D^{\diamond}$ and $N_R^{\diamond}$ are respectively the total number of local extrema  in the data and random catalogs. The lower part of Fig.  \ref{fig:onlypeaks}, indicates the peak pairs separated by $r$ in a magnified patch.

\subsection{General definition of bias factor}

 The general statistical expression for the relation between unweighted TPCF as the excess probability of finding a typical feature and the weighted TPCF which is known as auto-correlation function at the first-order approximation, reveals a linear and scale-independent bias factor. For a Gaussian random field, the excess probability of finding the sharp clipping pairs is statistically magnified by the fluctuation of random field  for long separation distance (angle) at high threshold, $\vartheta\gg 1$.   In this regime, we have $\Psi_{\rm pix-pix}(\theta;\vartheta)\sim e^{\mathcal{B}^2_{\rm pix}(\vartheta)C_{TT}(\theta)}-1$ with $\mathcal{B}_{\rm pix}(\vartheta)\sim \vartheta$ \citep{kaiser1984spatial,taqqu1977law,politzer1984relations,jensen1986n,Bardeen:1985tr,szalay88,szalay1988constraints1} (for extensive discussion see \cite[and references therein]{martinez2001statistics,desjacques2018large}).

 Now, we turn to the modulation of local maxima number density at threshold $\vartheta$ in the CMB map by the temperature fluctuations at the last scattering surface. In other words, we look for the relation between the unweighted TPCF of peaks and weighted TPCF of temperature fluctuations.  The general form of the bias factor enables us to estimate the unweighted TPCF of typical feature by using the weighted TPCF. Generally, we expect that the number density of peaks is enhanced where the temperature fluctuations are high.
For the CMB map, we define the peaks number density contrast as:
$\delta_{\rm pk} \equiv \frac{n_{\rm pk} - \langle n_{\rm pk} \rangle}{\langle
n_{\rm pk} \rangle }$. To determine the scale-independent bias factor averaged on all peak curvature values, we are
interested in examining a relation such as $\delta_{\rm pk} =
\mathcal{B}_{\rm pk}(\vartheta)\delta_T$. Following \cite{kaiser1984spatial} for sharp clipping statistics,  we
exploit a systematic relation between unweighted TPCF of local
extrema, $\Psi_{\diamond-\diamond}(\theta;\vartheta_1,\vartheta_2)$,  and
weighted TPCF of temperature fluctuations, $C_{TT}(\theta)$, for
separation angle, $\theta$, in analogy with $
\Psi_{\diamond-\diamond}(\theta;\vartheta)={\mathcal{B}}^2_{\diamond}(\vartheta)C_{TT}(\theta)$, when we ignore the scale-dependent part.
The relation mentioned is satisfied for large separation
angles. For very high threshold values, the pixel above threshold can
delineate the peak better than a small threshold and consequently we obtain $\mathcal{B}_{\rm pk}(\vartheta)\sim
\mathcal{B}_{\rm pix}(\vartheta)$ for large enough separation angles in  a Gaussian field
\citep{martinez2001statistics}.
To examine the scale-dependent part of the bias for peak statistics,  we rely on a more general model for bias in the Fourier space as: $\mathcal{B}_{\rm pk}(k, \vartheta)=\mathcal{B}_{\rm pk}(\vartheta)+\mathcal{B}_{k}^{\rm pk}(\vartheta)k^2$. Here $k$ represents the wavelength of the typical mode.  It turns out that  for either $\vartheta\gg 1$ or large scales, the value of scale-dependent part of bias is $\mathcal{B}_{k}^{\rm pk}(\vartheta)\to 0$ \citep{desjacques2010modeling,desjacques2018large}. We define
$\mathcal{B}^2_{\diamond}(\theta,\vartheta)\equiv\Psi_{\diamond-\diamond}(\theta;\vartheta)/C_{TT}(\theta)$ for proper range of $\theta$. Subsequently, any features existing in  $\mathcal{B}^2_{\diamond}(\theta,\vartheta)$ versus $\theta$ for a typical value of $\vartheta$ represent the contribution of the scale in this type of bias factor.  We will  assess the behaviour of such a bias factor and accordingly the consistency of {\it Planck} maps and associated state-of-the-art simulations.

\begin{figure*}
    \includegraphics[width=1.5\columnwidth]{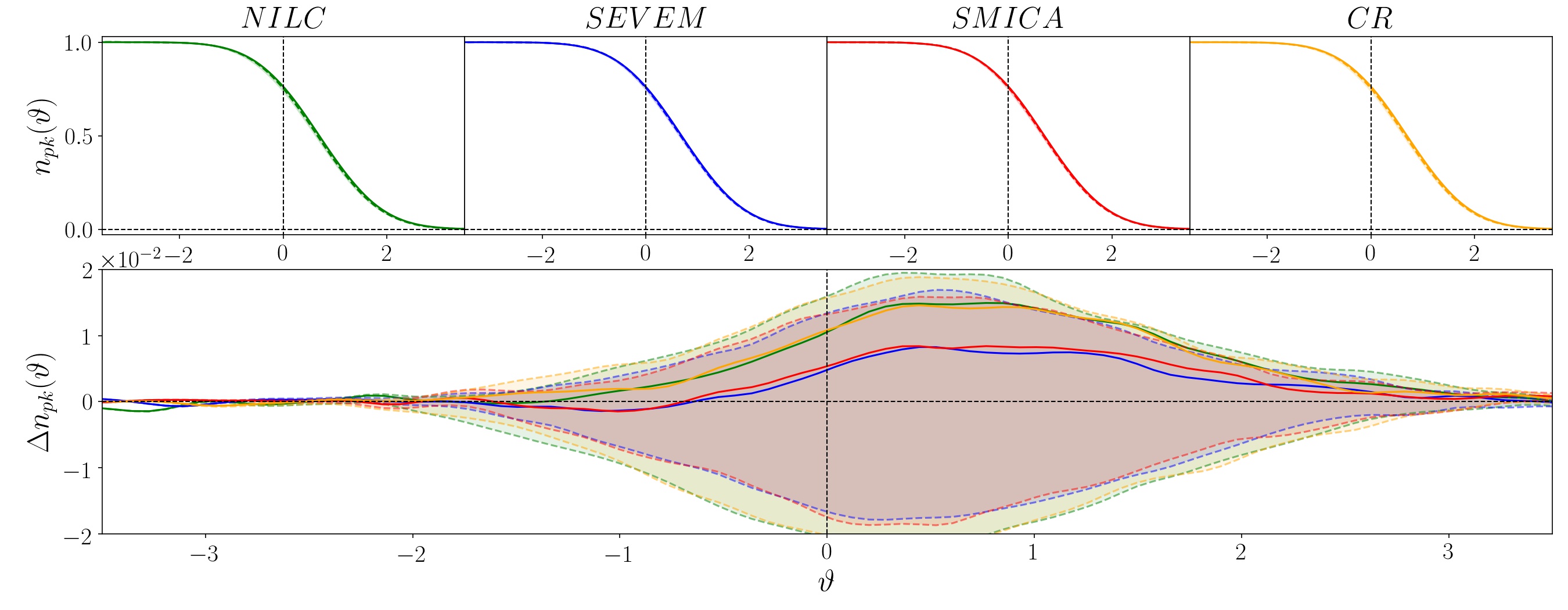}
    \includegraphics[width=1.5\columnwidth]{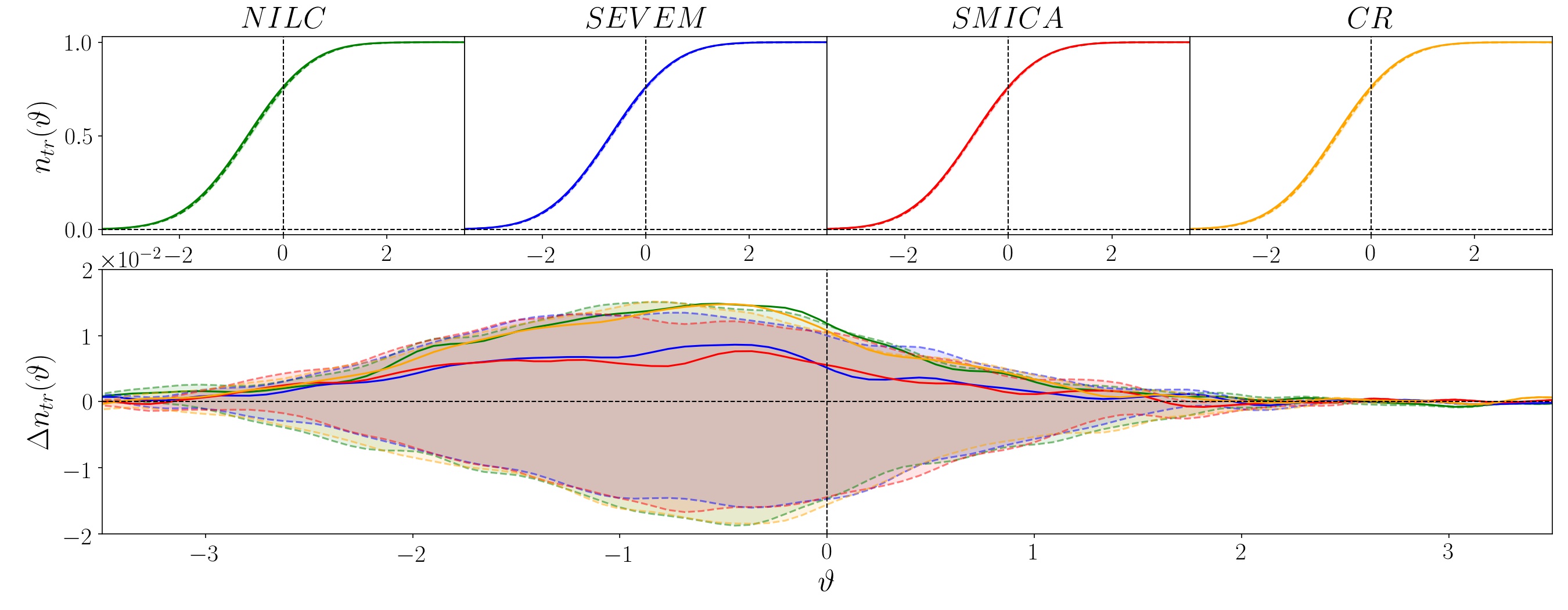}
    \caption{ Colour online:  Upper panel  corresponds to the cumulative number density of peaks above a threshold, $n_{\rm pk}(\vartheta)$,  for various observed CMB data (solid lines) compared to associated end-to-end simulations (dashed line) at the top part  and the corresponding differences, $\Delta n_{\rm pk}(\vartheta)$, at the bottom part of this panel, respectively. The  lower panel indicates the cumulative number density of troughs below a threshold, $n_{\rm tr}(\vartheta)$, at the top and the differences, $\Delta n_{\rm tr}(\vartheta)$, at the bottom, respectively. The difference plots are given to clarify how tiny the deviation is. The solid lines with different colours in mentioned plots show the residues of different component separations for observation and the shaded region corresponds to the $2\sigma$ confidence interval determined by the ensemble average on fiducial Gaussian CMB simulations. We considered $N_{\rm{side}}=2048$.}
    \label{fig:nextrema}
\end{figure*}

\section{Data description and simulation}\label{sec:data}
The {\it Planck} sky observational data sets contain full-sky maps at nine
frequency channels for temperature intensity. The polarization maps are available  only  for   30-353 GHz frequency range. Throughout this paper, we only focus on the temperature field. The mentioned data sets are processed by the $Planck$ team, resulting in different component
separation algorithms, namely \texttt{Commander-Ruler}
(\texttt{CR}), \texttt{NILC}, \texttt{SEVEM} and \texttt{SMICA}
\citep{akrami2020planckcomponent}; these are publicly available and are provided in the \texttt{HEALPIX}\footnote{\url{http://healpix.sourceforge.io}} format
\citep{Gorski:2004by}. Such component separation algorithms
allow us to achieve the largest possible sky area coverage. In
addition, these procedures can  remove galactic emission and
reconstruct the diffuse emission from our galaxy (see \cite
{dickinson2016cmb}  for a comprehensive description of the CMB
foreground). The latest proper mask and the corresponding fraction of
unmasked pixels used for the CMB data are specified by
$\texttt{UT78}$ and  $f_{\rm{sky}}=77.6 \%$, respectively
\citep{2016A&A...594A...9P,Ade:2015hxq}. To realize reliable statistical inferences
about the number density of the local extrema and associated
clustering, we also need fiducial simulations as the
reference sets and for debiasing in our statistical analysis.
To this end, we use 500 realizations of the end-to-end \footnote{\url{{pla.esac.esa.int}}} simulations \citep{plancke2elow,Aghanim:2018fcm}, where they are publicly available
on {\it Planck} Legacy
Archive\footnote{\url{https://pla.esac.esa.int/pla/\#maps}}. The
end-to-end fiducial CMB power spectrum is based on
the $\Lambda$CDM model  with the best-fitting {\it Planck} parameters \citep{Aghanim:2018eyx}. To interpret our results, the fiducial maps would be the Gaussian-based expectation. To avoid obtaining spurious results, we will compare the results in the context of local extrema analysis of each component separation outcomes with corresponding  simulations.

\section{Implementation on real data and synthetic data sets }\label{simul}
In this section, we apply our statistical measures based on critical sets explained in Section \ref{stat} on the {\it Planck} CMB maps and corresponding end-to-end simulated data sets.

\begin{figure*}
\begin{center}
\includegraphics[width=1.5\columnwidth]{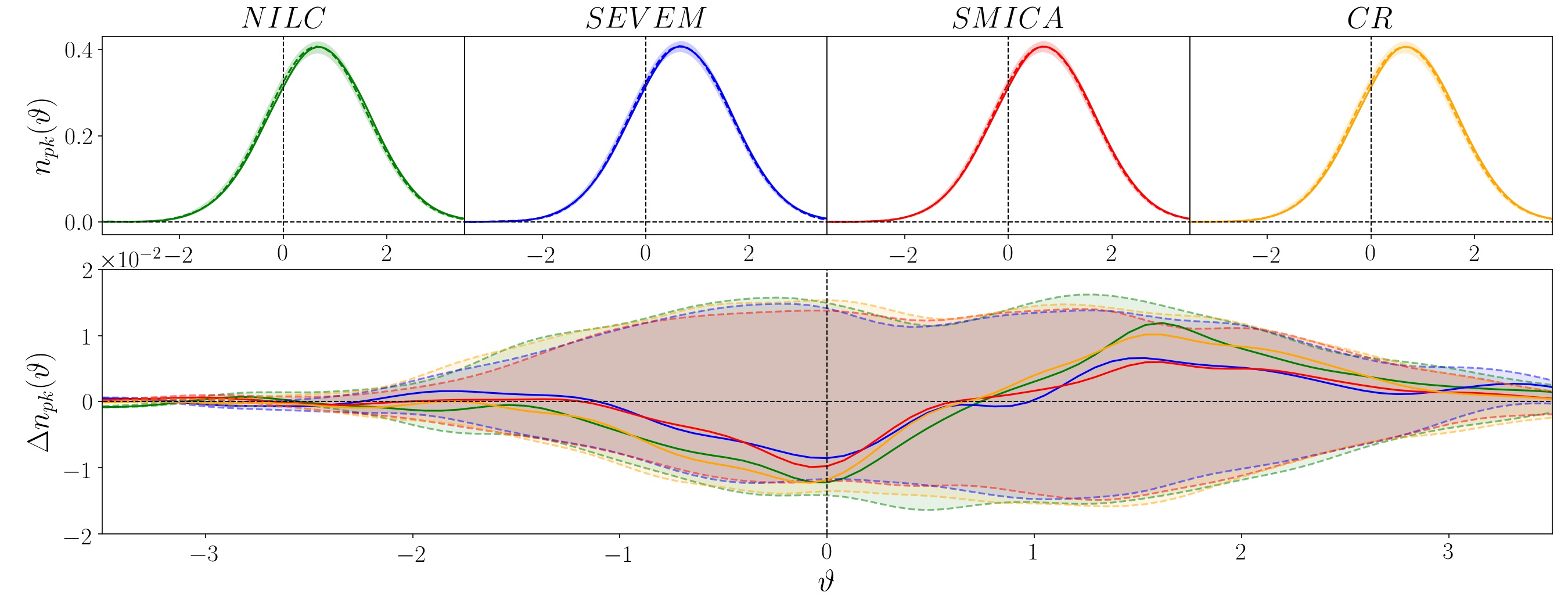}
\includegraphics[width=1.5\columnwidth]{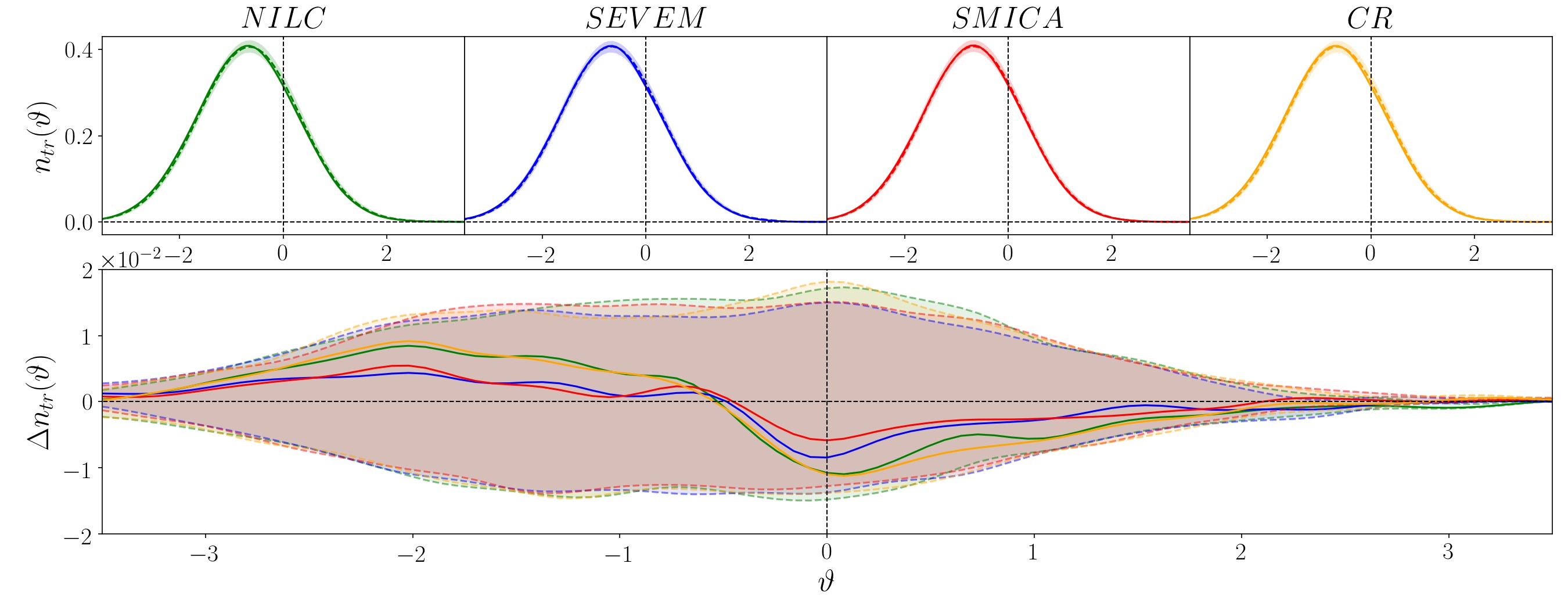}
\includegraphics[width=1.5\columnwidth]{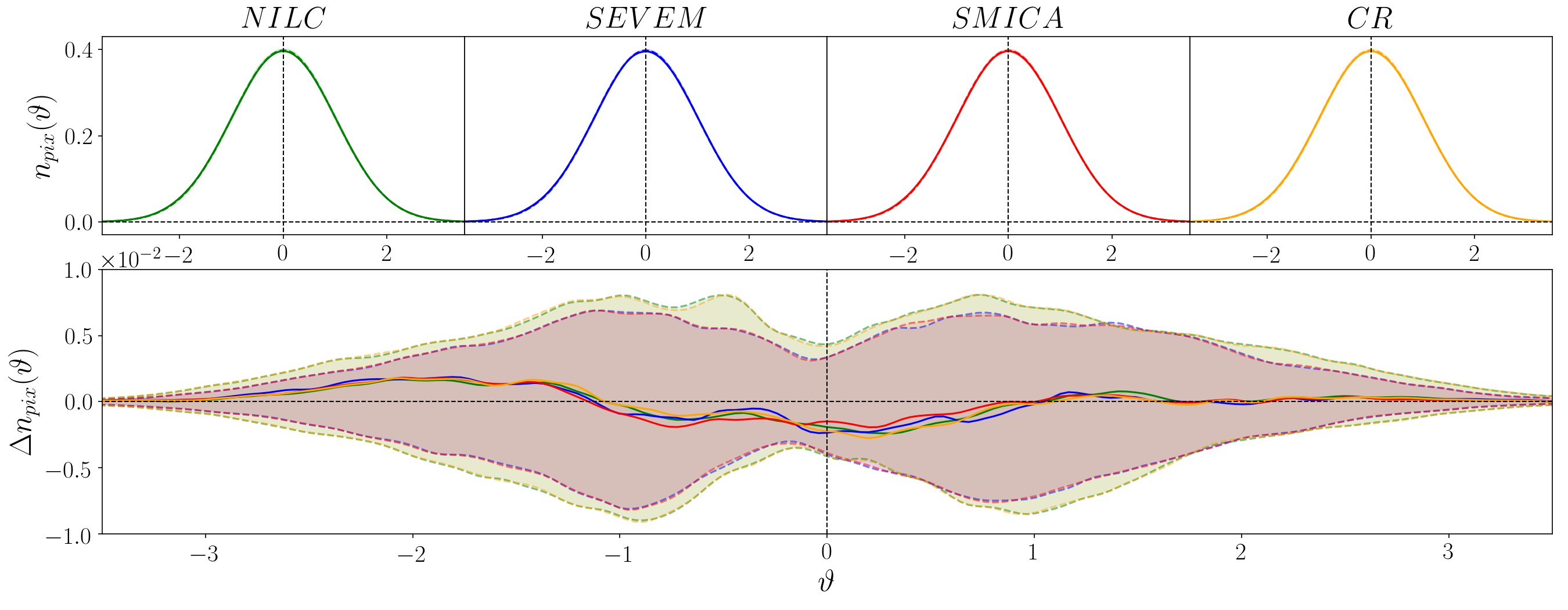}
\caption{The number density of peaks ({\it upper panel}) and troughs ({\it middle panel}) as a function of threshold for {\it Planck} data sets and corresponding Gaussian simulations. The number density of pixel at the threshold is illustrated in the lower panel.  In the lower part of each panel, we have computed the difference of number density with corresponding simulated map. The shaded region corresponds to the $2\sigma$ optimal variance error determined by fiducial Gaussian CMB map. The solid lines with different colours in the mentioned plots show the residues of different component separations for observation and the shaded region corresponds to the $2\sigma$ confidence interval determined by the ensemble average on fiducial Gaussian CMB maps. We considered $N_{\rm{side}}=2048$.}
\label{fig:numberdensity}
\end{center}
\end{figure*}

\begin{figure*}
\begin{center}
    \includegraphics[width=0.9\columnwidth]{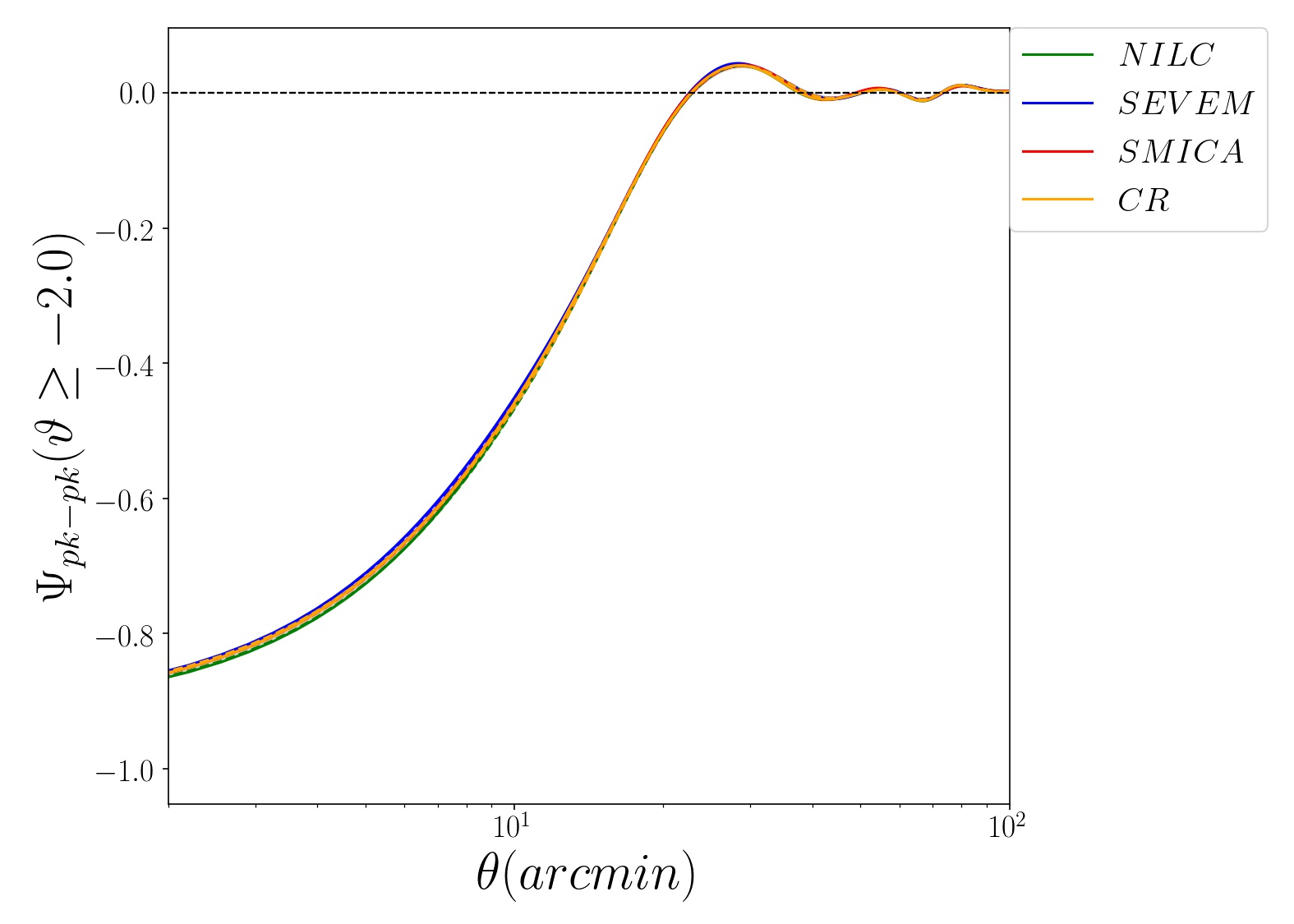}
    \includegraphics[width=0.9\columnwidth]{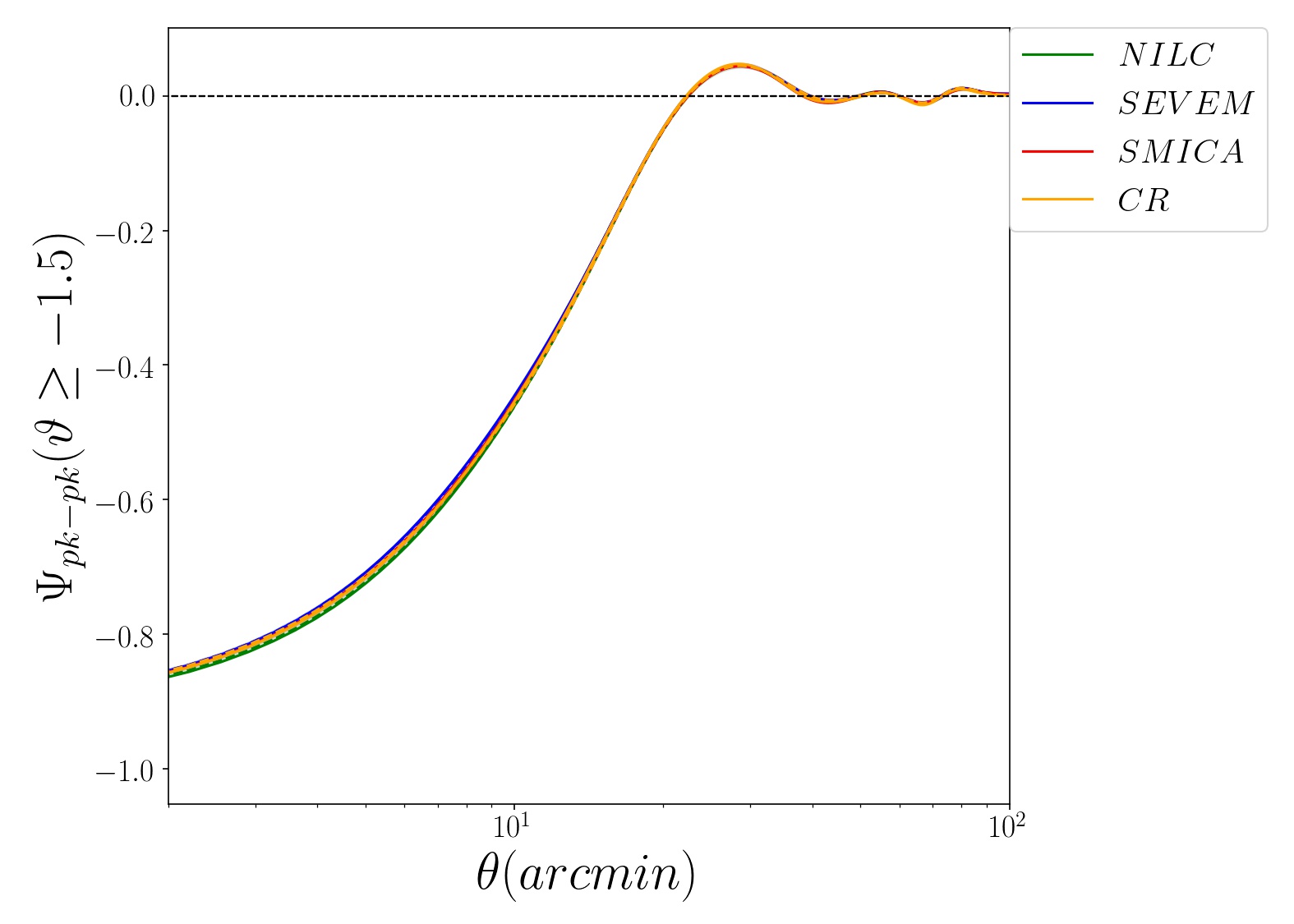}
    \includegraphics[width=0.9\columnwidth]{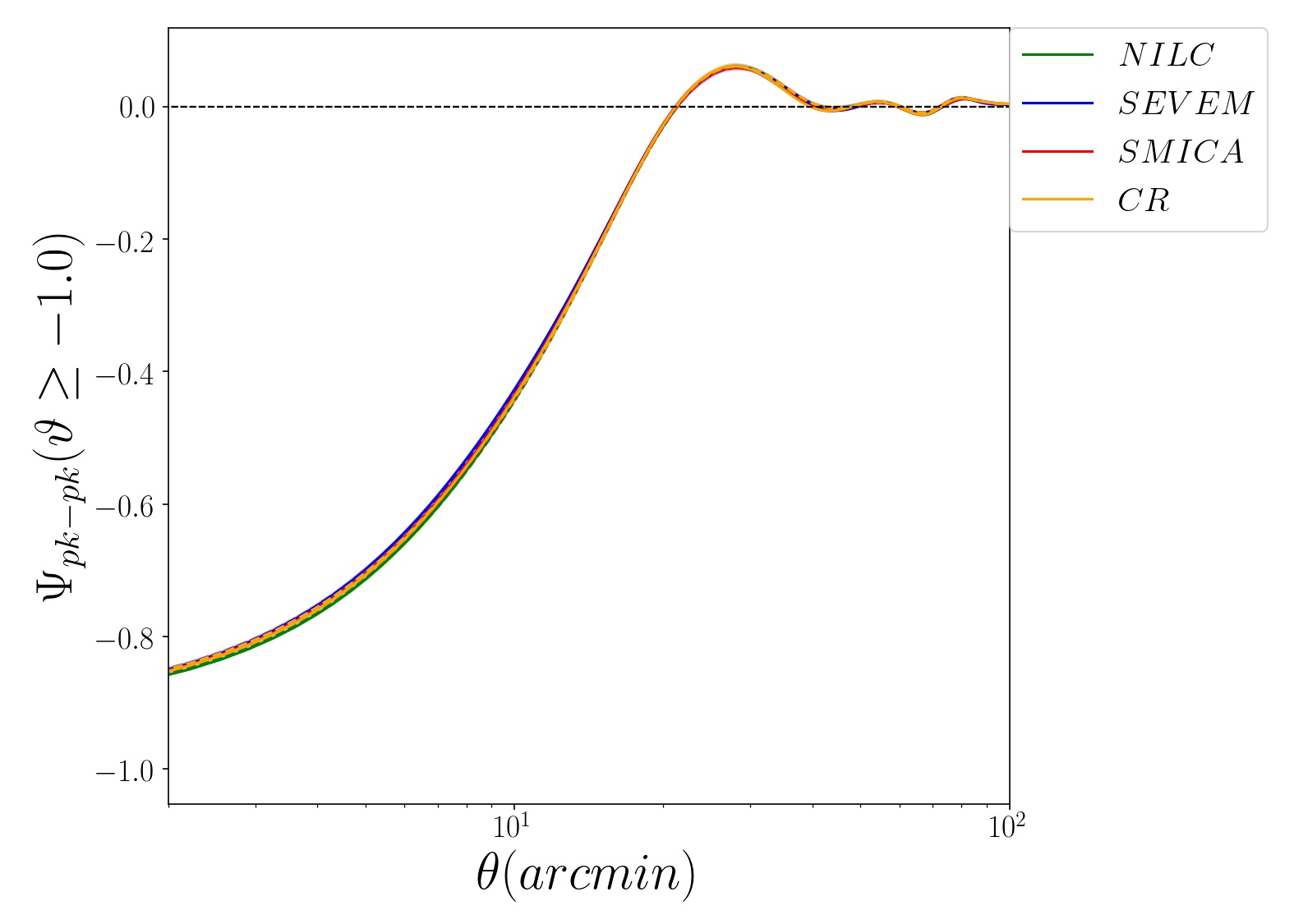}
    \includegraphics[width=0.9\columnwidth]{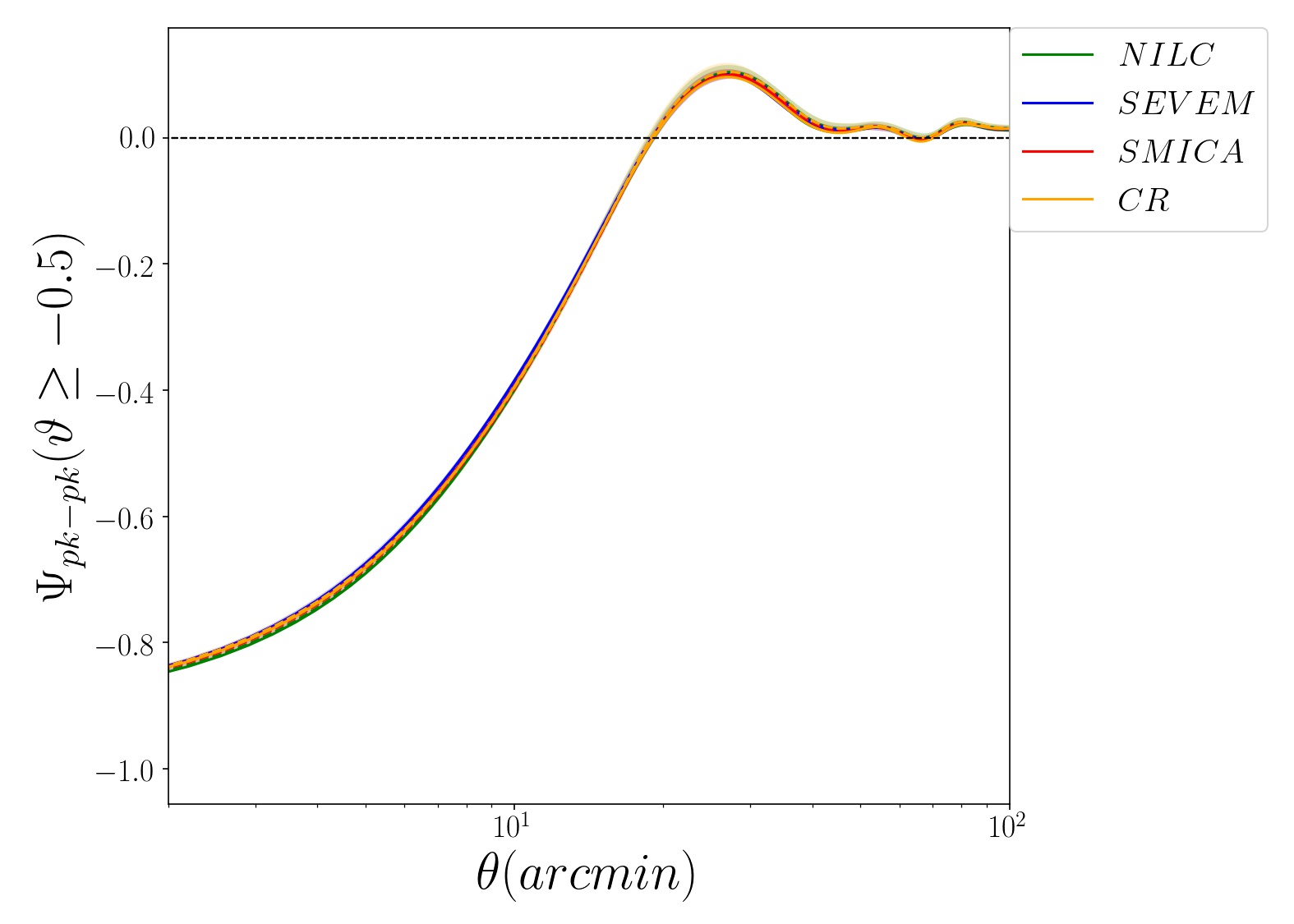}
    \includegraphics[width=0.9\columnwidth]{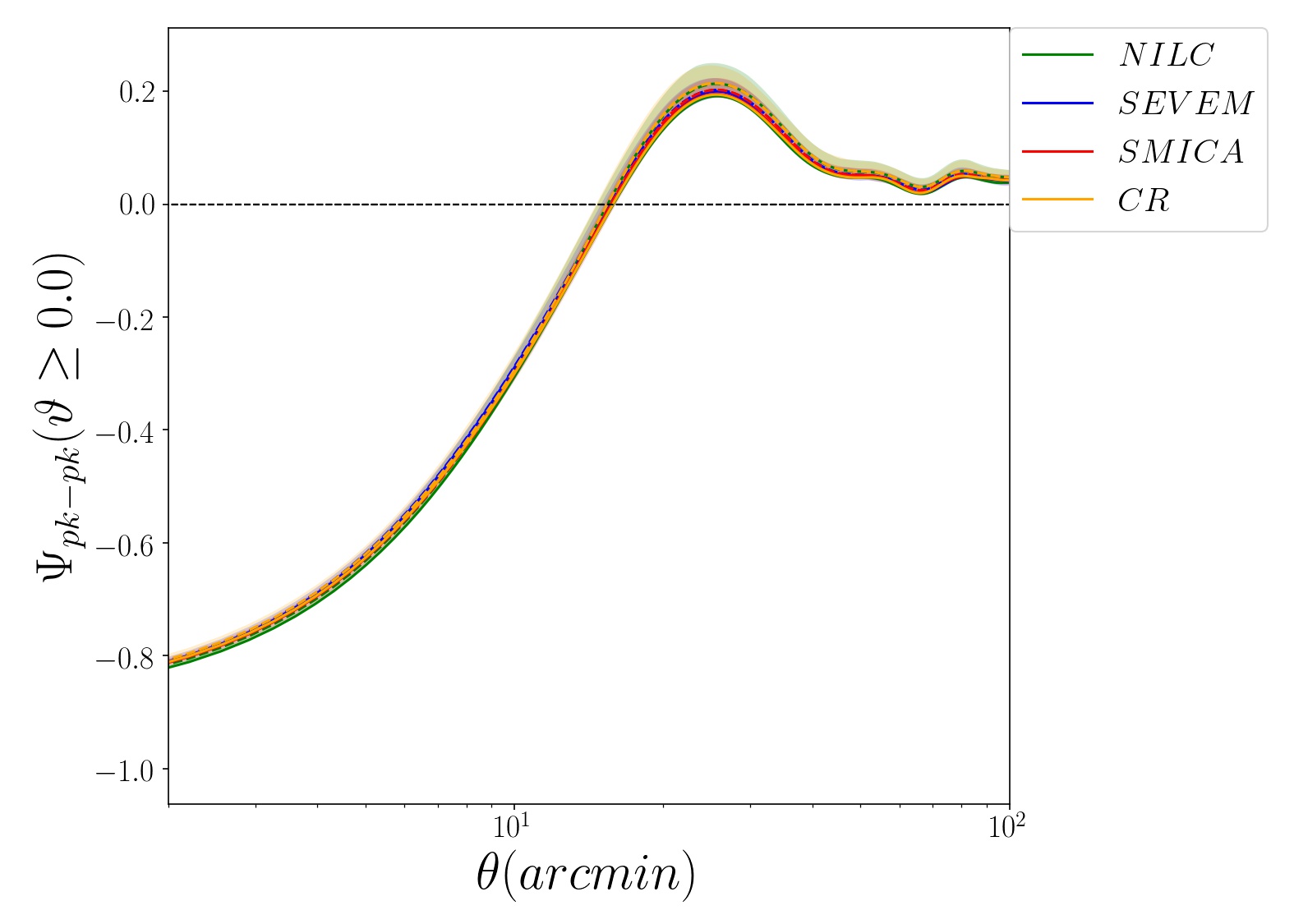}
    \includegraphics[width=0.9\columnwidth]{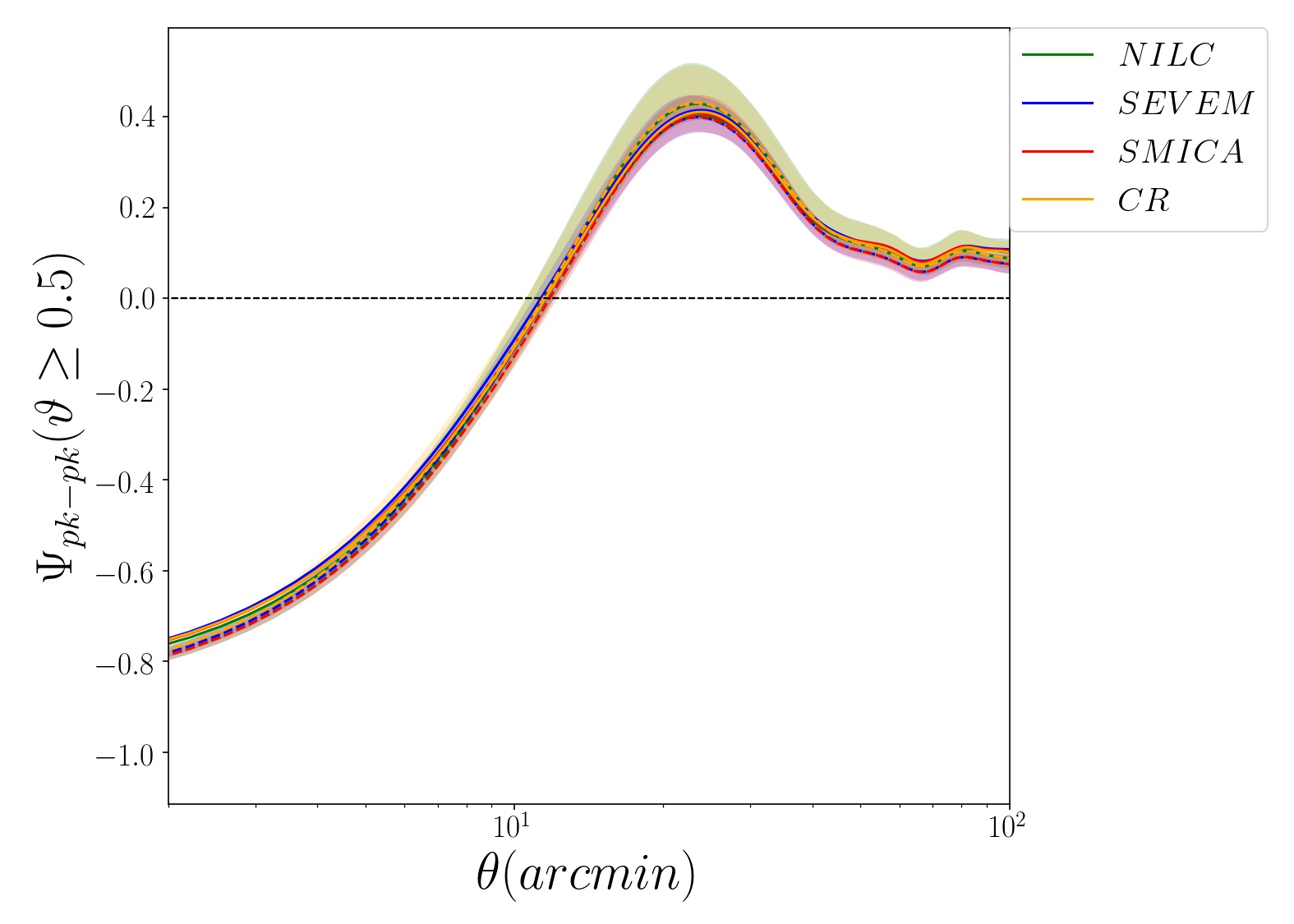}
    \includegraphics[width=0.9\columnwidth]{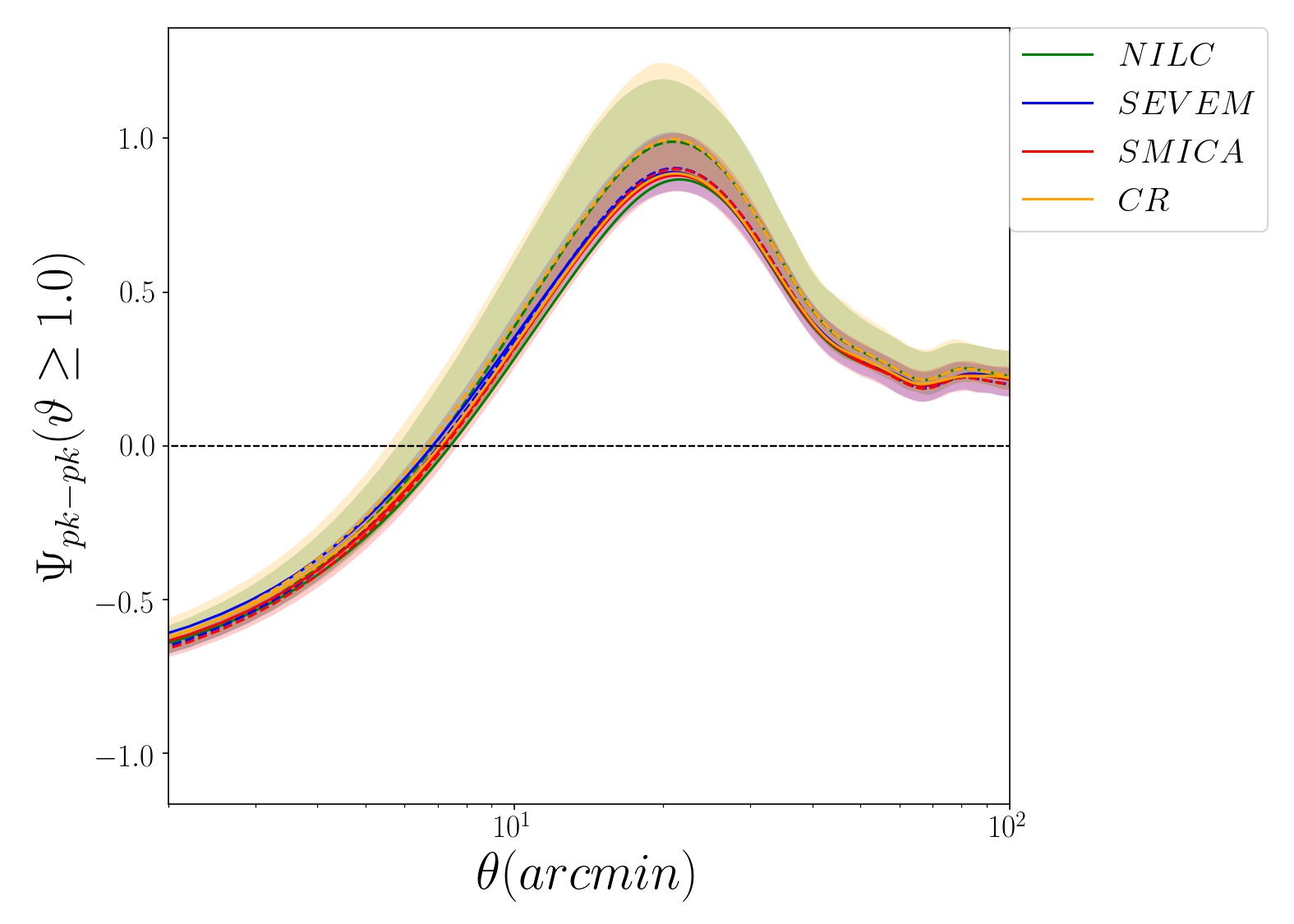}
    \includegraphics[width=0.9\columnwidth]{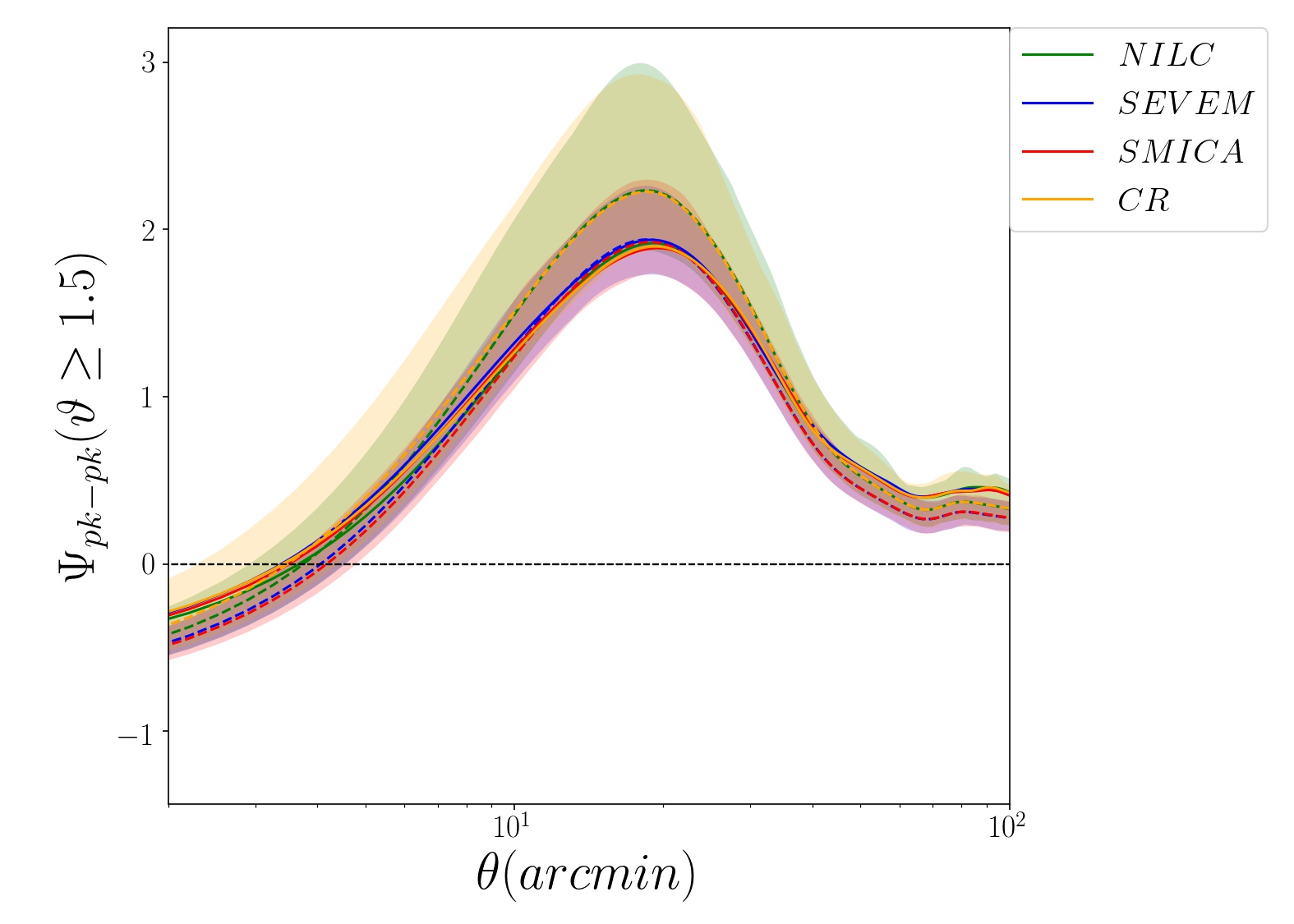}
    \caption{The unweighted TPCF of peaks for the CMB observed maps considering $\vartheta \ge [-2.0-1.5]$. The simulation in this plot corresponds to the end-to-end $\Lambda$CDM Gaussian CMB  map. The solid lines show the results for the different component separations, while  the dashed lines show ensemble average based on  end-to-end  simulations and the shaded region corresponds to the $2\sigma$ level of confidence.We considered $N_{\rm{side}}=2048$.}
    \label{fig:TPCFn}
    \end{center}
\end{figure*}

\subsection{Local extrema statistics of CMB map}
Critical regions among of excursion sets are efficient measures to recognize exotic features in the CMB maps. To start, we turn to one-point statistics of sharp clipping and local extrema.The latter is more  complicate due to the conditions required for extrema, while the former is more slightly less sensitive to searching for probable exotic features. 

Fig.  \ref{fig:nextrema} illustrates the normalized cumulative number density of local maxima for different types of observed data, while the lower panel illustrates the
cumulative number density of troughs below the threshold.  To recognize the deviation between each component separation product and associated end-to-end simulation, we compute $\Delta n_{\rm pk}$  and $\Delta n_{\rm tr}$ which are depicted in the lower part of upper and lower panels, respectively.  In mentioned plot, the solid lines correspond to the results for the observed map, while the dashed line are for associated end-to-end simulations. The shaded region represents the $2\sigma$ level of confidence. According to the $n_{\rm tr}$ statistics,  our results show that the \texttt{NILC} and \texttt{CR} have small deviations from corresponding simulations around the  $\sim \vartheta \in [-2-0]$, while the cumulative peak number density  of all observed maps are  consistent with each other. Our results also confirm that the values of the cumulative number density of peaks and troughs for all observed maps are higher than those values computed  for the wide range of thresholds.

Fig.  \ref{fig:numberdensity} depicts the number density of local maxima
(upper panel), local minima (middle panel) and pixel (lower panel) at the
threshold. All observations display very tiny deviation at mean threshold ($\vartheta \approx 0$).  We should point out that, according to the  local extrema number density at the mean threshold, all component separations data sets are located below  the corresponding values computed for simulations. It turns out that the sharp clipping statistics is less sensitive to non-Gaussianity and represents more consistency  with simulated maps (lower panel of Fig.   \ref{fig:numberdensity}). Another interesting
result is that, there is no significant  difference between
$n_{\rm pk}(\vartheta)$ and $n_{\rm tr}(-\vartheta)$ for all of the CMB maps, as
expected for a Gaussian field.


Going beyond the one-point statistics of critical sets provides a
proper opportunity for examining exotic features and enables us to reduce the
spurious and artificial effects due to undesired non-cosmological sources. Therefore, we use  unweighted TPCF of local extrema. To this end, we
check all the estimators introduced by Eqs.
(\ref{eq:pp-estimator1}),  (\ref{eq:pp-estimator2}) and
(\ref{eq:pp-estimator3}). Our analysis demonstrates that all results
are consistent with each other. Consequently, for the rest of this
paper, we will show what we have obtained by the  Hamilton estimator (Eq.
(\ref{eq:pp-estimator2})). We find all the peaks and troughs above
thresholds in the interval that corresponds to $ -3 \le\vartheta \le3$
with step size $0.5$ in each full sky simulated map and then apply
unweighted TPCF estimator to compute
$\Psi_{\diamond-\diamond}(\theta;\vartheta)$ for both observed and
simulated maps. The final results for simulation are given by doing ensemble average
over 500 realizations. Such results play the role of numerical
results for the Gaussian map.  The unweighted TPCF of peaks as a
function of separation angle is shown in Fig.  \ref{fig:TPCFn} for
{\it Planck} data above a given threshold for different component
separation algorithms. The solid lines are associated with {\it Planck} CMB maps, while the dashed lines represent the unweighted TPCF for end-to-end simulations. The clustering of local maxima  demonstrates that all observed map are consistent with the Gaussian hypothesis. However, there is a lack of peak clustering for high enough threshold around $10'\lesssim\theta\lesssim 30'$ compared to the {\it Planck} fiducial $\Lambda$CDM model. For $\vartheta\ge 1.0$, small difference between various component separations is also seen.

We also assess the unweighted TPCF of
peaks for $\delta_T<-\vartheta \sigma_0$ and for
$\delta_T>+\vartheta \sigma_0$. For an ideal Gaussian random field,
we expect to have $\Psi_{\diamond-\diamond}(\delta_T<-\vartheta
\sigma_0)=\Psi_{\diamond-\diamond}(\delta_T>+\vartheta \sigma_0)$
due to symmetry between peaks and troughs. Our results confirm that
observed CMB data are consistent with this expectation.  While
clustering of the pixels above and below a threshold for $WMAP$ data
done by \cite{Rossi:2009wm}, showed different results for large
separation angle . This achievement clarifies that peak-peak
statistics rather than pixel-pixel analysis is more robust in the
presence of probable un-resolve noises and other unknown effects. Also, the updated observed maps reveals more consistency with expected Gaussian properties, if we utilize the clustering of local extrema rather than pixels clustering.

\begin{figure}
\begin{center}
    \includegraphics[width=0.9\columnwidth]{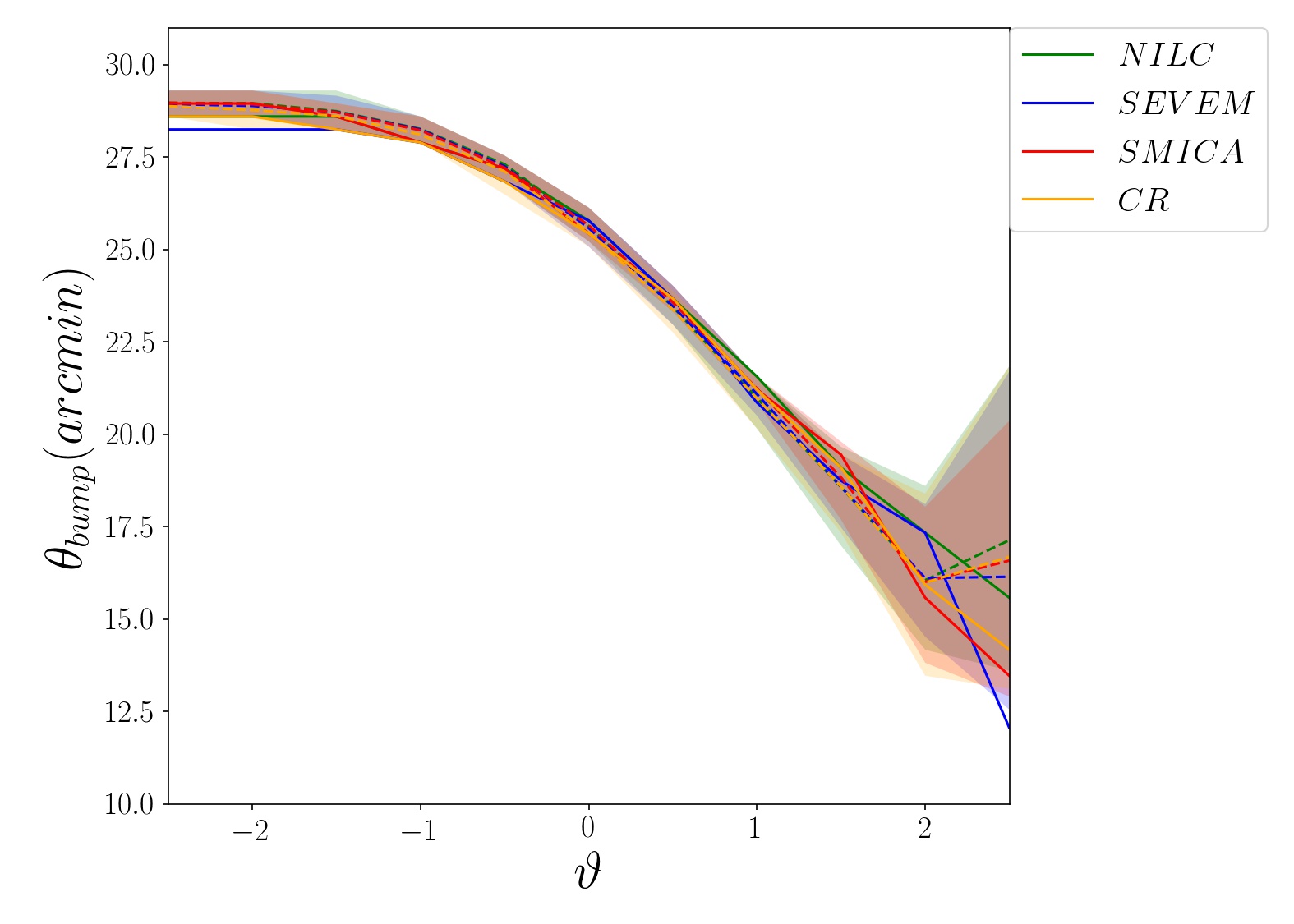}
    \includegraphics[width=0.9\columnwidth]{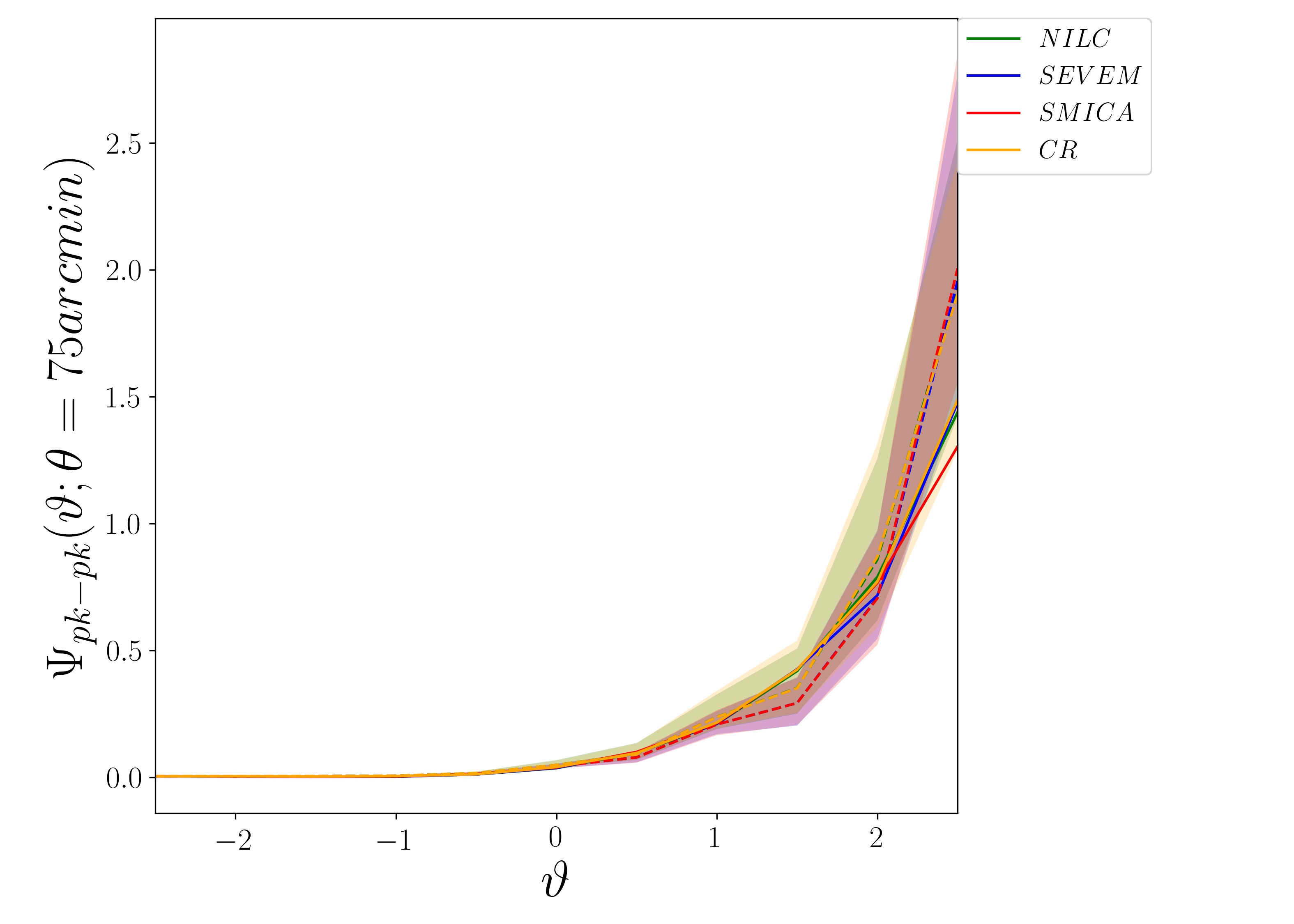}
    \caption{The value of angular separation at which the unweighted TPCF of peaks reaches its maximum value ($\theta_{\rm bump}$) as a function of threshold is indicated in the upper panel. The  lower panel shows the value of the unweighted TPCF of peaks around Doppler peak, $\theta\approx 75$ arcmin as a function of threshold. The solid lines show the results of different component separations for observation, the dashed lines correspond to the end-to-end simulations prepared for each observed map, and the shaded region corresponds to the $2\sigma$ region to see deviation from fiducial Gaussian CMB maps. We considered $N_{\rm{side}}=2048$.}
    \label{fig:bump_peak}
    \end{center}
\end{figure}
\begin{figure}
\includegraphics[width=0.93\columnwidth]{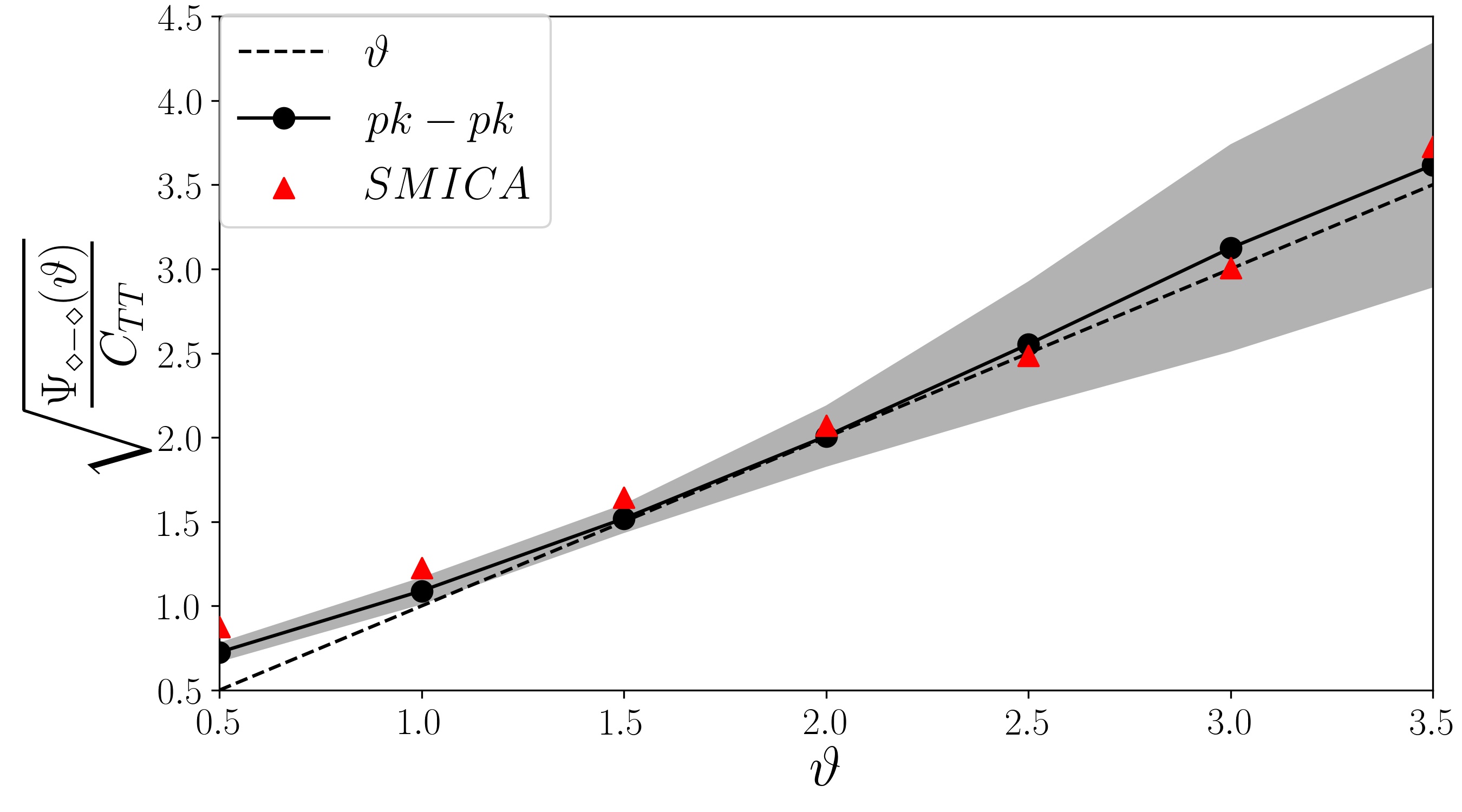}
\includegraphics[width=0.95\columnwidth]{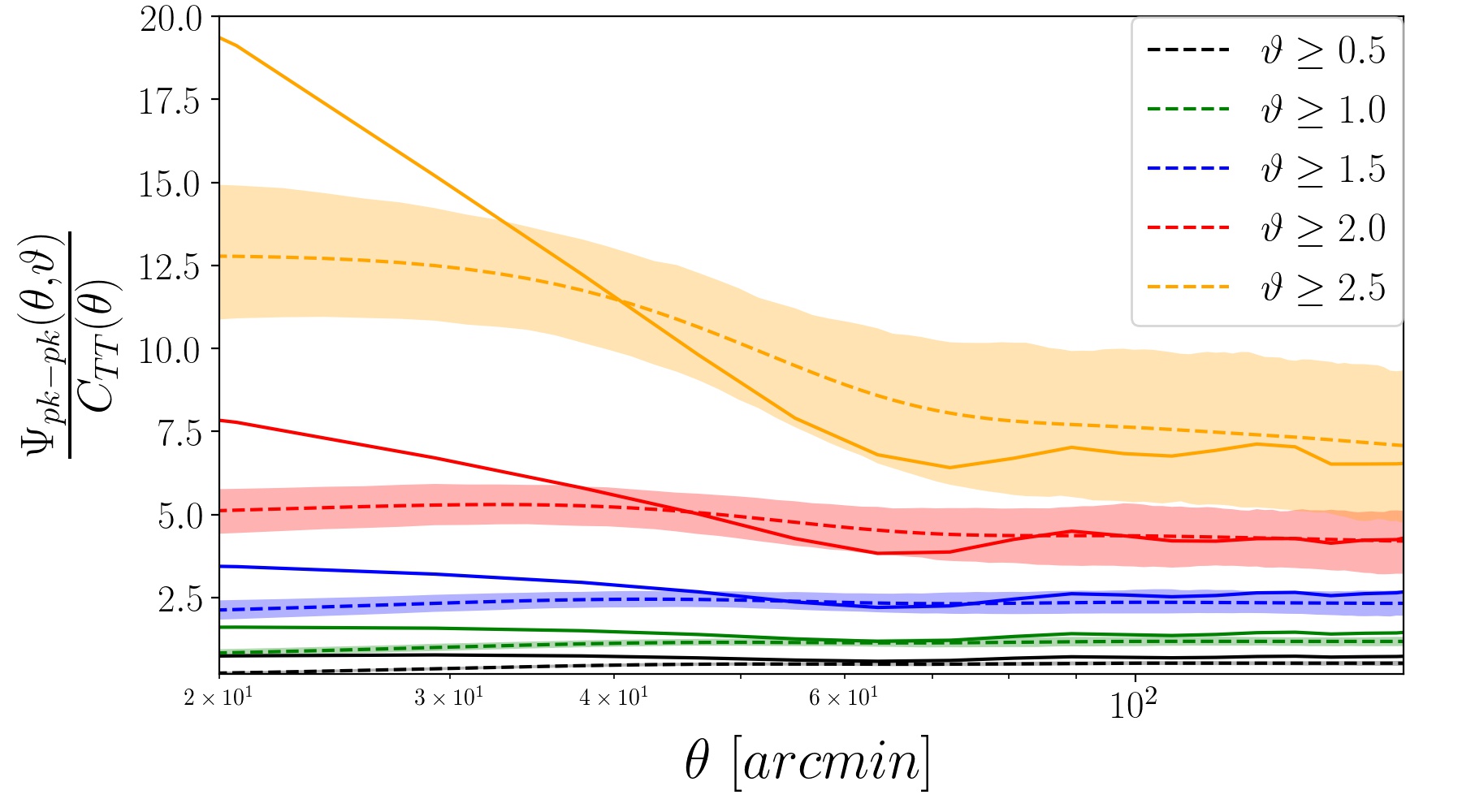}
\caption{The scale-independent bias factor for peaks above a threshold for CMB simulated data and \texttt{SMICA} maps is illustrated in the upper panel. The dashed line shows the results for pixels above a threshold given by theory for a Gaussian field while the filled circle symbol with a line is for peaks above threshold for a simulated CMB map. The symbols indicate the results for observation.  The lower panel corresponds to the evaluation of scale-dependent bias for peak statistics above a threshold.
The dashed  lines show results for the end-to-end simulated map corresponding to \texttt{SMICA} while the solid lines show for \texttt{SMICA}. The shaded region corresponds to the $2\sigma$ region to see deviation from the fiducial Gaussian CMB maps where $N_{\rm{side}}=2048$.  }
\label{fig:bias}
\end{figure}

\begin{table*}
\begin{center}
\caption{Probabilities of obtaining values of the $\chi^2$ statistics represented by $P(\chi^2>\chi^2_{\diamond-\diamond}(\vartheta))$, $P(\chi^2>\chi^2_{\rm pix})$, $P(\chi^2>\chi^2_{\rm pk})$ and $P(\chi^2>\chi^2_{\rm tr})$ for the {\it Planck} fiducial $\Lambda$CDM Gaussian model at least as large as the observed values for various observed maps.  Here we consider $N_{\rm side}=2048$.}
\label{tab:datRedng}
\begin{tabular}{@{}cccccccccc}
\hline \hline
Map/Measure &  $n_{pix}$ &  $n_{pk}$ &  $n_{tr}$ &  $\Psi_{tr-tr}(\vartheta \leq 0.0)$ &  $\Psi_{tr-tr}(\vartheta \leq -1.0)$ &  $\Psi_{pk-pk}(\vartheta \geq 0.0)$ &  $\Psi_{pk-pk}(\vartheta \geq 1.0)$ \\
\hline
\texttt{NILC} & 0.42 &      0.88 &      0.57 &                               0.13 &                                0.74 &                               0.22 &                               0.70 \\
\hline
\texttt{SEVEM}&0.88 &      0.54 &      0.72 &                               0.36 &                                0.86 &                               0.46 &                               0.79 \\
\hline
\texttt{SMICA}& 0.60 &      0.22 &      0.64 &                               0.62 &                                0.26 &                               0.21 &                               0.40 \\
\hline
\texttt{CR}   &0.26 &      0.12 &      0.72 &                               0.59 &                                0.62 &                               0.37 &                               0.54 \\
\hline \hline
\end{tabular}
\end{center}
\end{table*}

The position of so-called bump for unweighted TPCF of peaks above a given threshold versus $\vartheta$ is indicated in the upper panel of Fig.  \ref{fig:bump_peak}. The lower panel corresponds to the value of unweighted TPCF of peaks around Doppler peak, $\theta\sim 75$ arcmin, as a function of threshold.  In this plot, we depict the results for observed maps by the solid lines and  dashed lines are refereed  to the associated end-to-end simulations drawn in the same colours.  According to the semi-analytical approach (see \cite{Heavens:1999cq}), by increasing the threshold, the unweighted TPCF of peaks in a Gaussian CMB map reaches  its maximum value for lower separation angle. Such behaviour can be justified by means of the distribution of local maxima at higher threshold which is more distinguishable from random catalog. Applying the different beam size can also wash out the peaks and suppressing the bump in $\Psi$ for all thresholds \citep{Heavens:1999cq}. The value of $\theta_{\rm bump}$ for observed maps is almost less than synthetic data for high thresholds.  The value of extrema clustering around the Doppler peak for higher threshold in the {\it Planck} observed data is less than the corresponding value in the simulated maps. However, taking into account statistical errors, mentioned deficiency is not statistically significant.

According to the researches done by \cite{takada2000gravitational,takada2001detectability}, the bump and main trough around the Doppler peak in the unweighted TPCF of peaks are more sensitive to  the effect of gravitational lensing on the CMB photons which are randomly deflected by foreground ranging from large scale structures to cosmic strings networks. In other words, redistribution of peak in the CMB map from intrinsic separation and distribution by weak lensing phenomenon have unique signature on the $\Psi_{\diamond-\diamond}$.  The contribution of weak lensing on the $\Psi_{\diamond-\diamond}$ leads to mitigate the depth of Doppler peak and even suppressing the maximum value of $\Psi$ as a function of separation angle \citep{takada2001detectability}. To put proper constraint on the amplitude of the mass fluctuations, we need more deep troughs around $\theta\approx70'-75'$, consequently, according to Fig.  \ref{fig:TPCFn}, we should compute $\Psi_{\diamond-\diamond}$ for mean threshold due to more compatibility between different separation components of CMB achieved for higher thresholds around Doppler peak.

The upper panel of Fig.  \ref{fig:bias} indicates the scale-independent bias factor for peaks above a threshold for \texttt{SMICA} map and corresponding Gaussian simulated data sets. Here, we average on the ratio of $\Psi_{\rm pk-pk}/C_{TT}$  for large enough separation angle for each threshold denoted by symbols.  For a Gaussian stochastic field, we expect to have ${\mathcal{B}}_{\rm pix}(\vartheta)\sim \vartheta$, for high threshold \citep{kaiser1984spatial,taqqu1977law,politzer1984relations,jensen1986n,Bardeen:1985tr,szalay88,szalay88}. Semi-analytical investigation at very high threshold clarifies that, peak and pixel statistics become identical and our results are compatible with Gaussianity. Such result have been obtained for other type of observed maps and we avoid to show them for convenience.

To examine the contribution of angular scale in the bias, we compute ${\mathcal{B}}_{pk}(\theta,\vartheta)$ as a function of $\theta$ for some typical thresholds. The lower panel of Fig.  \ref{fig:bias} illustrates the unweighted TPCF of peaks to the temperature fluctuations correlation function  ratio, versus scale for various thresholds. In this panel the dashed lines correspond to the Gaussian CMB simulated map extracted from end-to-end pipeline associated with \texttt{SMICA}, while solid line is devoted  to \texttt{SMICA} map. The presence of some features particularly in the angular scale interval, $20$  arcmin $\lesssim\theta \lesssim80$ arcmin, confirms that for small scale, the footprint of scale essentially appears in the bias factor, while for large enough separation angle, such features are diminished. In the latter regime and for $\vartheta\gg1$, we expect to have a plateau for $\mathcal{B}^2$ and its value is proportional to $\vartheta^2$. It is worth mentioning that, the results for $Planck$ data sets are compatible with that expected for Gaussian field, while for small scales, we have some deviations from the results given for simulated map.  By increasing the threshold, the consistency between $Planck$ data and simulations increases. Other component separations reveal almost similar results.

\begin{figure*}
\includegraphics[width=1.7\columnwidth]{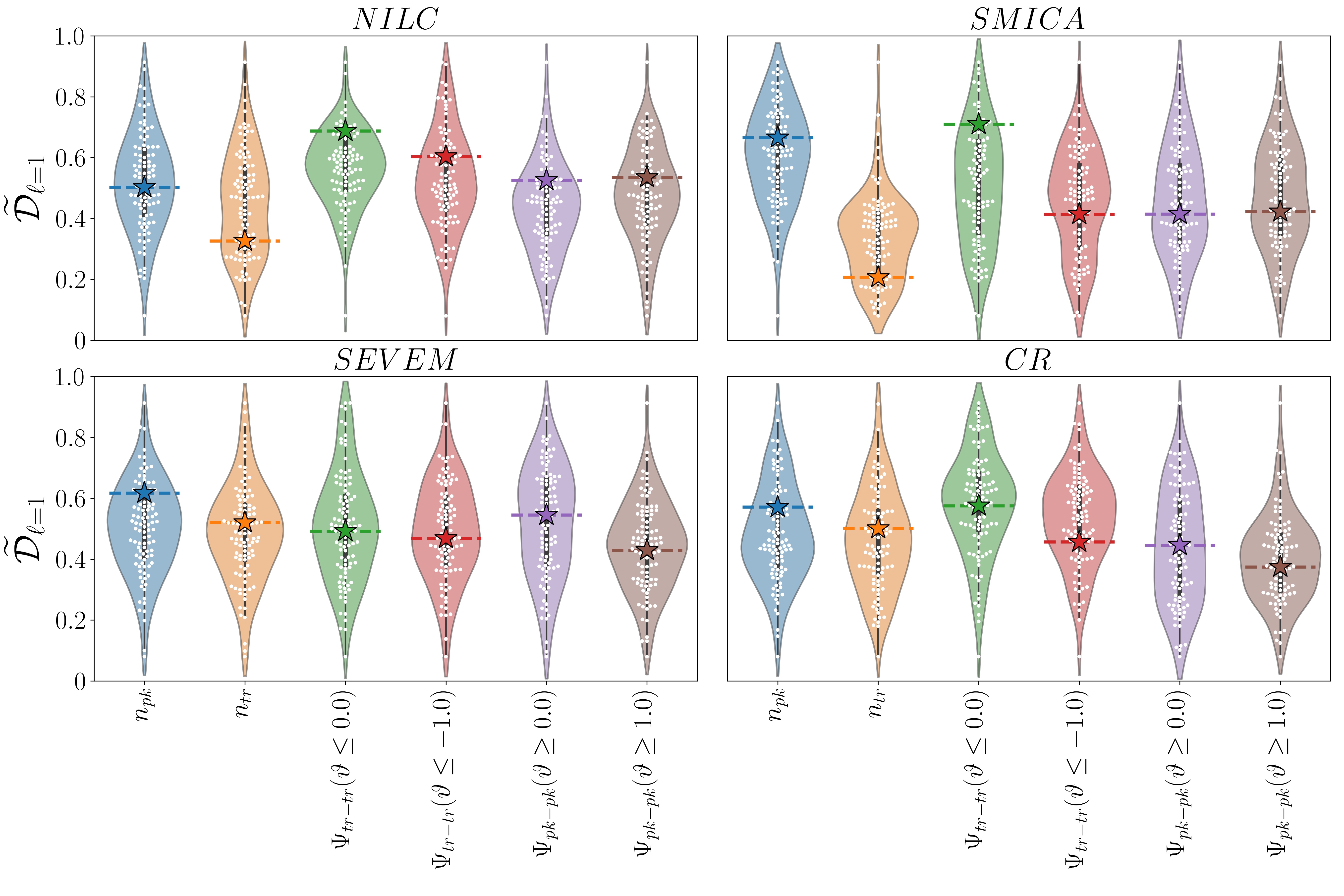}
\caption{The violin plot for amplitude of the dipole power spectrum of local $\chi^2$ (Eqs. (\ref{eq:chipktr}) and  (\ref{eq:chiasym2})) considering the $n_{\rm pk}$, $n_{\rm tr}$ marginalized on all available thresholds, $\Psi_{\rm pk-pk}$ and $\Psi_{\rm tr-tr}$ integrated out on some separation angles. We took  threshold ranges, $\vartheta\ge0.0$ and $\vartheta\ge1.0$ for clustering of peaks, while, for the unweighted-TPCF of troughs, the following threshold ranges have been taken into account,  $\vartheta\le-1.0$, $\vartheta\le0.0$. The star symbol in each panel indicates the value computed for the observed map. }
\label{fig:powerchi}
\end{figure*}

\subsection{Testing Gaussian Hypothesis}
Non-Gaussianity in the CMB data can be produced by late phenomena such as gravitational lensing, the Sunyaev-Zel'dovich effect,  contaminations from foreground and residual point sources. Primordial non-Gaussianity can be generated by a sequence of phase transitions and deviation from uncorrelated initial fluctuations during the inflationary epoch \citep{komatsu2002pursuit,Lewis:2006fu,ade2014plancknon,2016A&A...594A..17P,akrami2020plancknong}.

To probe the hypothesis of Gaussianity, many statistical approaches have been proposed \citep[and references therein]{Heavens:1999cq,Tojeiro:2005mt,Rossi:2009wm,Rossi:2010hu,ade2014plancknon,2016A&A...594A..17P,novaes2016local,cole2018persistent,buchert2017model,planck2013results,Ade:2015hxq,akrami2020planck}. The {\it Planck} team also carried out the peak statistics in addition to the other methods to asses non-Gaussianity in $Planck$ maps  \citep{Ade:2015hxq,akrami2020planck} (for WMAP data set \citep{larson2004hot,Larson:2005vb,2009MNRAS.396.1273H}). Here, we implement clustering of local extrema in addition to number density of peaks, troughs and sharp clipping to evaluate the Gaussianity of the {\it Planck} CMB maps.  In our analysis, we utilize 500 realizations of the end-to-end simulation pipeline maps \citep{Aghanim:2018fcm}.  For a Gaussian stochastic field, $\mathcal{P}(\mathcal{A}_{\mu1};\mathcal{A}_{\mu2})$ is modeled by the Gaussian multivariate function. Any deviation from Gaussianity is encoded in a joint PDF leading to a deviation in the unweighted TPCF from a typical Gaussian model. Therefore, this quantity is a powerful measure for testing the Gaussian  hypothesis \citep{Heavens:1999cq,Rossi:2009wm}.

Based on an efficient estimator for local extrema clustering introduced by Eq. (\ref{eq:pp-estimator2})\footnote{We have also carried out other estimators represented by Eqs. (\ref{eq:pp-estimator2}) and (\ref{eq:pp-estimator3}) for some of our analysis  and  we concluded that different measures give consistent results.}, we compute $\Psi_{\diamond-\diamond}^{\rm sim}(\theta;\vartheta)$ for a given threshold, $\vartheta$, for  simulated maps. Here the symbol $\diamond$ can be replaced by "pk" for local maxima above a given threshold or "tr" for local minima below a given threshold.  Each computed $\Psi_{\diamond-\diamond}^{\rm sim}(\theta;\vartheta)$ for a given threshold is divided into 20 classes for separation angle in the range of $\theta\in [5-100 {\rm arcmin}]$ and finally we record corresponding values in an array.
Relying on computed results for $\Psi_{\diamond-\diamond}^{\rm sim}(\theta;\vartheta)$, we make covariance matrix, ${\mathcal{C}}_{\diamond-\diamond}$, with size $20\times20$.  In order to determine the significance of deviation from the Gaussian hypothesis, the following  chi-square is computed for both  {\it Planck} data sets and associated end-to-end simulated maps as:
\begin{eqnarray}\label{eq:chiTPCFnongaussian}
\chi^2_{\diamond-\diamond}(\vartheta,i)&=&\sum_{m=1}^{20}\sum_{n=1}^{20}[ \Psi_{\diamond-\diamond}(\theta_m;\vartheta,i)-\langle\Psi^{\rm sim}_{\diamond-\diamond}(\theta_m;\vartheta,i,j)\rangle_j]\nonumber\\
&&\times{\mathcal{C}}^{-1}_{\diamond-\diamond}(\theta_m,\theta_n;\vartheta,i)\nonumber\\
&&\times[\Psi_{\diamond-\diamond}(\theta_n;\vartheta,i)-\langle\Psi^{\rm sim}_{\diamond-\diamond}(\theta_n;\vartheta,i,j)\rangle_j]
\end{eqnarray}
where $j=1,...,500$ for a simulated map associated with any type of observed CMB and $i=1,...,4$ for different component separations, namely $\texttt{CR}$, $\texttt{NILC}$, $\texttt{SMICA}$ and $\texttt{SEVEM}$ maps.  Also ${\mathcal{C}}_{\diamond-\diamond}(\theta_m,\theta_n;\vartheta,i)\equiv\langle [\Psi^{\rm sim}_{\diamond-\diamond}(\theta_m;\vartheta,i,j)-\langle\Psi^{\rm sim}_{\diamond-\diamond}(\theta_m;\vartheta,i,j)\rangle_j] [\Psi^{\rm sim}_{\diamond-\diamond}(\theta_n;\vartheta,i,j)-\langle\Psi^{\rm sim}_{\diamond-\diamond}(\theta_n,\vartheta,i,j)\rangle_j]\rangle_j$. We also compute the probability density function of $\chi^2_{\diamond-\diamond}(\vartheta,i)$ for a Gaussian simulated map, $P(\chi^2_{\diamond-\diamond}(\vartheta))$. Accordingly, for each observed data sets, we compute $P(\chi^2>\chi^2_{\diamond-\diamond}(\vartheta))$,  namely, the probability of obtaining values for the $\chi^2_{\diamond-\diamond}(\vartheta,i)$  statistics (Eq. (\ref{eq:chiTPCFnongaussian})) for associated simulations  at least as large as those given from the observed maps.  Table \ref{tab:datRedng} shows $P(\chi^2>\chi^2_{\diamond-\diamond}(\vartheta))$ for clustering of peaks and troughs for different thresholds. It is worth noting that, according to the measure of one-point statistics for critical sets, namely, number density of sharp clipping, peaks, and trough, we also evaluate  the non-Gaussianity according to  the proper $\chi^2$ defined for the measures mentioned. The covariance matrix for one-point statistics reads as $ {\mathcal{C}}_{\diamond}(\vartheta,\vartheta',i)\equiv\langle [n^{\rm sim}_{\diamond}(\vartheta,i,j)-\langle n^{\rm sim}_{\diamond}(\vartheta,i,j)\rangle_j] [n^{\rm sim}_{\diamond}(\vartheta',i,j)-\langle n^{\rm sim}_{\diamond}(\vartheta',i,j)\rangle_j]\rangle_j$.  The results for examining the Gaussianity hypothesis according to $n_{\rm pix}$, $n_{\rm pk}$ and $n_{\rm tr}$ integrated out on all available thresholds  are reported in Table \ref{tab:datRedng}. Our results in the context of the TPCF of peaks, number density of local extrema, and pixels, demonstrate that, there is no significance deviation from the Gaussian hypothesis.

\begin{table*}
\begin{center}
\caption{ The probability of obtaining dipole amplitude for the  fiducial map at least as large as the value given from  observed map, based on  the local maxima and minima for one- and two-point statistics.}
\label{tab:ass1}
\begin{tabular}{@{}ccccccccc}
\hline \hline
 Map/Measure  & $n_{pk}$ &  $n_{tr}$ &  $\Psi_{tr-tr}(\vartheta \leq 0.0)$ &  $\Psi_{tr-tr}(\vartheta \leq -1.0)$ &  $\Psi_{pk-pk}(\vartheta \geq 0.0)$ &  $\Psi_{pk-pk}(\vartheta \geq 1.0)$ \\
\hline
\texttt{NILC}  &       0.54 &      0.70 &                                 0.16 &                                  0.31 &                                 0.24 &                                 0.36 \\
\hline
\texttt{SEVEM} &  0.25 &      0.40 &                                 0.54 &                                  0.53 &                                 0.50 &                                 0.56 \\
\hline
\texttt{SMICA} & 0.38 &      0.77 &                                 0.15 &                                  0.64 &                                 0.56 &                                 0.58 \\
\hline
\texttt{CR}    &  0.27 &      0.40 &                                 0.58 &                                  0.74 &                                 0.48 &                                 0.57 \\
\hline
\hline
\end{tabular}
\end{center}
\end{table*}

\subsection{Asymmetry in clustering of local extrema}

A great deal of research  has  been done to establish robust measures and to examine the statistical isotropy of the CMB in both intensity and polarization \citep[and references therein]{hajian2003measuring,hajian2006testing,copi2004multipole,copi2006large,hanson2009estimators,rath2013testing,rath2013direction,akrami2014power,planck2013results,Ade:2015hxq,akrami2020planck,rath2018testing,adhikari2018statistical,akrami2020plancknong}. The well-known cosmological model  supports the hypothesis of statistical homogeneity and isotropy of initial conditions, essentially leading to CMB fluctuations behaving as an isotropic random field when the secondary anisotropies are well removed. Subsequently, testing such fundamental characteristics plays an important role in  examining  the standard cosmological scenario. Meanwhile, the presence of some anomalies such as power asymmetry and deviation from statistical isotropy in a range of multipoles have been reported \citep[and references therein]{eriksen2004asymmetries,prunet2005constraints,hansen2009power,hanson2009estimators,hoftuft2009increasing,rath2013testing,rath2013direction,akrami2014power,planck2013results,Ade:2015hxq,rath2018testing,akrami2020planck}.

 Here, we rely on the local extrema statistics and corresponding clustering  to evaluate the probable asymmetry superimposed on the observational maps.
 We calculate $n_{\diamond}$ and $\Psi_{\diamond-\diamond}$ for all unmasked patches size $7.5^{\circ}\times 7.5^{\circ}$ centered on the  pixels of a \texttt{HEALPIX} $N_{\rm side} = 8$  \citep{Gorski:2004by}.  This is a proper patch size for  searching  for asymmetry \citep{akrami2014power}. Therefore, we denote  our measures for each direction in each simulation as: $n_{\diamond}(\vartheta,i,q,j)$ and $\Psi_{\diamond-\diamond}(\theta_m;\vartheta,i,q,j)$ in which  $m=1,...,20$ corresponds to the label of separation angle bin, $q=1,...,768$ represents the number of non-overlapped patches, and $j=1,...,500$ indicates the label of simulated map, while, for observed map, we have $i=1,...,4$ corresponding to the different component separation maps (\texttt{CR}, \texttt{NILC}, \texttt{SMICA}, \texttt{SEVEM}).
Based on the number density of local extrema, we compute $\langle n^{\rm sim}_{\diamond}(\vartheta,i,q,j)\rangle_j$. The significance of the difference between $n_{\diamond}(\vartheta,i,q, j)$ for each patch in each observed map can be derived by $\chi^2_{\diamond}(i,q)$ for $q$th patch and $i$th map as follows:
\begin{eqnarray}\label{eq:chipktr}
\chi^2_{\diamond}(i,q)&=&\sum_{\vartheta,\vartheta'}[ n_{\diamond}(\vartheta,i,q)-\langle n^{\rm sim}_{\diamond}(\vartheta;i,q,j)\rangle_{j}]\nonumber\\
&&\times{\mathcal{C}}^{-1}_{\diamond}(\vartheta,\vartheta',i,q)[n_{\diamond}(\vartheta',i,q)-\langle n^{\rm sim}_{\diamond}(\vartheta',i,q,j)\rangle_{j}]\nonumber\\
\end{eqnarray}
where ${\mathcal{C}}_{\diamond}(\vartheta,\vartheta',i,q)\equiv\langle [n^{\rm sim}_{\diamond}(\vartheta,i,q,j)-\langle n^{\rm sim}_{\diamond}(\vartheta,i,q,j)\rangle_j] [n^{\rm sim}_{\diamond}(\vartheta',i,q,j)-\langle n^{\rm sim}_{\diamond}(\vartheta',i,q,j)\rangle_j]\rangle_{j} $ for each direction.  Using the computed $\chi^2$ for one-point statistics of local extrema,  we construct  the associated sky map for each observed maps as well as corresponding simulations. Then, we compute the angular power spectrum for local $\chi_{\diamond}^2$ maps. The amplitude of dipole for local $\chi_{\rm pk}^2$ and  $\chi_{\rm tr}^2$ defined by Eq. (\ref{eq:chipktr}) is represented by $\mathcal{D}^{\diamond}_{\ell=1}$ for the {\it Planck} maps. For convenience, we normalized  computed dipole amplitude to the range of $\widetilde{\mathcal{D}}^{\diamond}_{\ell=1}\in[0,1]$, as illustrated in Fig.  \ref{fig:powerchi}. In this violin plot, the `star' symbols are associated with observed maps and scattered dots correspond to the simulations. According to the value of angular power spectrum for local $\chi^2_{\diamond}$ at $\ell=1$, we compute  the probability of obtaining values for the $\mathcal{D}_{\ell=1}^{\diamond}$ for associated simulations  at least as large as those given from the observed maps $P(\mathcal{D}^{\diamond}_{\ell=1}>\mathcal{D}^{\diamond,{\rm obs.}}_{\ell=1})$ reported  in Table \ref{tab:ass1}.
For the unweighted TPCF of local extrema, we also define a local $\chi^2_{\diamond-\diamond}(\vartheta,i,q)$ for threshold $\vartheta$, $q$th patch and $i$th map as:
\begin{eqnarray}\label{eq:chiasym2}
\chi^2_{\diamond-\diamond}(\vartheta,i,q)&=&\sum_{m,n}[ \Psi_{\diamond-\diamond}(\theta_m;\vartheta,i,q)-\langle\Psi^{\rm sim}_{\diamond-\diamond}(\theta_m;\vartheta,i,q,j)\rangle_j]\nonumber\\
&&\times{\mathcal{C}}^{-1}_{\diamond-\diamond}(\theta_m,\theta_n;\vartheta,i,q)\nonumber\\
&&\times[\Psi_{\diamond-\diamond}(\theta_n;\vartheta,i,q,j)-\langle\Psi^{\rm sim}_{\diamond-\diamond}(\theta_n;\vartheta,i,q,j)\rangle_j]\nonumber\\
\end{eqnarray}
where ${\mathcal{C}}_{\diamond-\diamond,mn}(\vartheta,i,q)\equiv\langle [\Psi^{\rm sim}_{\diamond-\diamond}(\theta_m;\vartheta,i,q,j)-\langle\Psi^{\rm sim}_{\diamond-\diamond}(\theta_m;\vartheta,i,q,j)\rangle_j] [\Psi^{\rm sim}_{\diamond-\diamond}(\theta_n;\vartheta,i,q,k)-\langle\Psi^{\rm sim}_{\diamond-\diamond}(\theta_n;\vartheta,i,q,j)\rangle_k]\rangle_{jk} $ for each threshold and each direction. The local $\chi^2_{\diamond}(\vartheta,i,q)$ for $q$th patch in $i$th map considering different types of measures has been computed. Accordingly, for each measure, a sky map for $\chi_{\diamond}^2$ is constructed. Now, we are able to calculate the angular power  spectrum for each map. The amplitude power spectra for $\ell=1$ considering  different statistical measures are illustrated in Fig.  \ref{fig:powerchi}. The results  for \texttt{NILC}, \texttt{SMICA}, \texttt{SEVEM} and \texttt{CR} have been compared to the results determined for associated simulations in Fig.  \ref{fig:powerchi}. Table \ref{tab:ass1} reports the probability of obtaining dipole amplitudes for fiducial map at least as the value given for observed map. Our results show that there is no significant deviation from isotropic field when we use either one- or two-point  peak/trough statistics. It is worth noting that, \texttt{SMICA} has  different behaviour compared to other observed maps, when we consider clustering of the local trough for $\vartheta \le 0.0$.

Our statistical measures  can be used for identifying probable anomalous patches according to the procedure done by \cite{novaes2016local,marques2018isotropy}.

\subsection{Cosmic String Network Detection}

A series of phase transitions could  have happened in the very early Universe and meanwhile depending on the topology of the potential of the underlying field, we expect to obtain point-like (mono-pole), line-like (cosmic string (CS)), and the texture of topological defects due to spontaneous symmetry breaking  in the expanding and cooling Universe \citep{Kibble:1976sj,Kibble:1980mv22,Hindmarsh:1994re,Vilenkin:2000jqa,Copeland:2009ga,Polchinski:2004hb}. 

In particular, the cosmic string (CS) network is predicted to exist by hybrid inflation, brane-word and superstring theories  \citep{Kibble:1976sj,Zeldovich:1980gh,Vilenkin:1981iu,Vachaspati:1984dz,Vilenkin:1984ib,Shellard:1987bv,Hindmarsh:1994re,Vilenkin:2000jqa,Sakellariadou:2006qs,Bevis:2007gh,Depies:2009im,Bevis:2010gj,Copeland:1994vg,Sakellariadou:1997zt,Sarangi:2002yt,Copeland:2003bj,Pogosian:2003mz,Majumdar:2002hy,Dvali:2003zj,Kibble:2004hq,HenryTye:2006uv}.  The energy-density characterization of CS is given by the string tension: $\frac{G\mu}{c^2}=\order{\frac{\varpi^2}{M_{\rm Planck}^2}}$. Here $M_{\rm Planck}\equiv\sqrt{\hbar c/G}$ is the Planck mass, $c$ indicates the speed of light, and $\varpi$ is the energy of symmetry-breaking scale. The search for the footprint of CS network leads to the discovery of proper bounds on $G\mu$ (see \cite{Ade:2013xla,vafaei2017multiscale,sadr2018cosmic} and references therein).

To find the upper bound on the CS tension in the {\it Planck} data  using the local extrema clustering approach, we follow same recipe for the simulation CS-induced CMB map as discussed by \cite{Bennett:1990, Ringeval:2005kr,Fraisse:2007nu,vafaei2017multiscale,sadr2018cosmic}. We use high-resolution flat-sky $60^2$ deg$^2$ patches of CMB maps extended of a full sky simulation for large redshift interval by map stacking method \citep{Bouchet:1988hh, Ringeval:2012tk}.
The CS tensions used in this work are in the range $2.6\times10^{-11} \leq G\mu \leq 5.0\times10^{-7}$ classified into 18 classes for each simulation category.  The simulated map for a given $G\mu$ is constructed as $T=T_{\rm Gaussian}+G\mu T_{\rm String}+N$, where $T_{\rm Gaussian}$, $T_{\rm String}$ and $N$ respectively correspond to the end-to-end map associated with each component separation, the normalized string simulated map and  the proper noise component.  The proper beam effect has been taken into account  \citep{Ade:2013xla}.  We compare $\Psi_{\rm pk-pk}(\vartheta\ge 0)$ computed for different observations and the one computed for various $G\mu$ simulations.

For a given $G\mu$, we compute the $\Psi_{\rm pk-pk}(\vartheta\ge 0)$ for different observations and for various $G\mu$ simulations. The covariance matrix is also considered as: $\mathcal{C}_{\diamond-\diamond}(\theta_m,\theta_n;G\mu,\vartheta,i)\equiv\langle [\Psi_{\diamond-\diamond}^{\rm sim}(\theta_m;G\mu, \vartheta,i,j)-\langle \Psi_{\diamond-\diamond}^{\rm sim}(\theta_m;G\mu, \vartheta,i,j)\rangle_j ]
[\Psi_{\diamond-\diamond}^{\rm sim}(\theta_n;G\mu, \vartheta,i,j)-\langle \Psi_{\diamond-\diamond}^{\rm sim}(\theta_n;G\mu, \vartheta,i,j)\rangle_j ]\rangle_j$, and $\chi^2_{\diamond-\diamond}(G\mu,\vartheta,i)$. Now for the observed map, we also determine corresponding  $\chi^2_{\diamond-\diamond}(\vartheta,i)$. Finally we compare the observation and simulations by checking the inequality as $P_{{\rm C.L.}}\ge \int_{\chi^2>\chi^2_{\rm pk-pk}(\vartheta)}^{\infty}P(\chi^2(G\mu,\vartheta,i))d\chi^2(G\mu,\vartheta,i)$. The $P_{{\rm C.L.}}$ is adopted for a given confidence level (C.L.).
Therefore, the minimum value of $G\mu$ for which the mentioned inequality is satisfied will be considered by the upper value of CS tension recognized in the observations. We report the $G\mu^{\rm (up)}$  for $\vartheta\ge 0.0$ in Table \ref{tab:cosmic} at $95.5\%$ level of confidence. Comparing our upper bound on CS tension with that of reported by $Planck$ team confirms that taking into account the clustering local extrema achieves almost consistent upper bound determined by considering bispectrum and Minkowski functionals \citep{Ade:2013xla}.

\begin{table}
\begin{center}
\caption{ The upper bound on the tension of cosmic-strings network, $G\mu^{(\rm up)}$, utilizing $\Psi_{\rm pk-pk}(\vartheta\ge 0)$ as a criterion for recognition.}\label{tab:cosmic}
\begin{tabular}{@{}cc}
\hline \hline
 Map &  $G\mu^{(\rm up)}$ (95.5\%)   \\
\hline
\texttt{NILC}  &  $8.38 \times 10^{-7}$ \\
\hline
\texttt{SEVEM} &  $6.71 \times 10^{-7}$ \\
\hline
\texttt{SMICA} &  $5.59 \times 10^{-7}$ \\
\hline
\texttt{CR}    &  $7.17 \times 10^{-7}$ \\
\hline
\hline
\end{tabular}
\end{center}
\end{table}

\section{Summary and conclusions}
The CMB map as a (1+2)-D stochastic field includes thermodynamic temperature and two types of polarization. The mentioned components contain useful information ranging from the early epoch to the late time era. Various  physical phenomena have different footprints on the CMB map. The stochastic nature of CMB fluctuations motivates us to rely on geometrical and topological measures to achieve  deep insight through  the physical mechanisms and associated evolutions. In this paper, we have focused on thermodynamic temperature fluctuations and  we addressed the critical sets properties not only in one-point statistics but also in a two-points analysis. After a comprehensive exploration of  different research, we turned to the  robust perturbative approach to determine the joint PDF of different components of the CMB random field to clarify some examples of excursion and critical sets in the form of one- and two-point statistics. In particular, we derived the perturbative definition of the number density of pixels above a threshold up to $\mathcal{O}(\sigma_0^3)$.  By means of the excess probability of finding a typical feature, we computed the unweighted TPCF of local extrema and revisited the semi-analytical definition of the unweighted TPCF of peaks and troughs. In practice, utilizing a semi-analytical approach may encounter the finite-size effect. Therefore we considered three powerful estimators for the rest part of our analysis (Eqs. (\ref{eq:pp-estimator1}), (\ref{eq:pp-estimator2}) and (\ref{eq:pp-estimator3})).

In order to prepare a robust framework for comparison of the Gaussian prediction and those  computed for observed maps, we applied reliable estimators for unweighted TPCF on the {\it Planck} fiducial $\Lambda$CDM model \citep{plancke2elow,Aghanim:2018fcm} and various component separations as observed by {\it Planck} \citep{akrami2020planckcomponent}.

One-point statistics of local extrema and sharp clipping in the form of probability density as a function of threshold and also the cumulative number density function have been computed for both end-to-end simulations and observed  {\it Planck} temperature CMB maps. Based on the normalized cumulative number density of troughs, we found that  \texttt{NILC} and \texttt{CR} represent very tiny deviations from the corresponding simulations around the  $\vartheta \in [-2-0]$, while the cumulative number density of peaks confirms the consistency for all component separations. Numbers of peaks, troughs, and sharp clippings for different observed maps compared to the corresponding end-to-end simulated data sets  indicated similar results, supporting the Gaussianity hypotheses in the different separation components. However, the pixel statistics at the threshold was less sensitive to the non-Gaussianity. As depicted in Fig. \ref{fig:nextrema}, the mean value of the cumulative number density of peaks and troughs for the observed maps is almost higher than the simulations. In addition, the value of the number density of local extrema at $\vartheta \approx 0$ for all observed maps is less than expected in the Gaussian simulation (see Fig.  \ref{fig:numberdensity}).

The unweighted TPCF of local extrema illustrated consistent
behaviour in different separation components in all thresholds. For
high enough thresholds, the $\Psi_{\rm pk-pk}$ was almost less than
the value expected for the fiducial Gaussian simulations. However,
the clustering of local extrema reveals good consistency between
different component separations (see Fig. \ref{fig:TPCFn}). Our
results revealed that foreground and shot-noise have been excluded
in a proper way from different component separations and there is
good consistency between the end-to-end simulations and
corresponding observed maps when we are dealing with clustering of
local extrema. The symmetry behaviour for
$\Psi_{\diamond-\diamond}(\delta_T\ge \vartheta \sigma_0)$ and
$\Psi_{\diamond-\diamond} (\delta_T\le \vartheta \sigma_0)$ was
confirmed when we considered peaks and troughs rather than pixels
reported by \cite{Rossi:2009wm} for {\it WMAP}.  The value of
$\theta$ around the  Doppler peak as a function of threshold was
decreasing (Fig. \ref{fig:bump_peak}). Our results demonstrated
that, for $\vartheta\sim0$, we are able to put robust constraint on
the amplitude of the mass function according to the value of
$\Psi_{\diamond-\diamond}(\theta\approx 70-75)$ arcmin. The
scale-independent bias factor for peaks above threshold at high
threshold demonstrated $\mathcal{B}_{\rm pk}\sim \vartheta$ which is
compatible for sharp clipping in the Gaussian regime, for {\it
Planck} sets.  The scale-dependent part of the bias for peak
statistics illustrated some features  in the angular scale interval,
$20{\rm arcmin}\lesssim\theta \lesssim80{\rm arcmin}$. For small
scales, some deviations between CMB simulated maps and {\it Planck}
data have been recognized. The higher value of  threshold implies
the better consistency between $\mathcal{B}_{\rm pk}$ for observed
and simulated CMB maps (Fig. \ref{fig:bias}).

To give quantitative evaluation of non-Gaussianity, we defined the proper $\chi^2$ and computed the probability of finding the sample with $\chi^2\ge\chi^{2}_{\diamond-\diamond}$ (Table \ref{tab:datRedng}). Accordingly, we found that all of statistical measures considered in this paper supported  the Gaussian hypothesis.

Asymmetry in the context of $n_{\diamond}$ and $\Psi_{_{\diamond-\diamond}}(\theta;\vartheta)$ for patch of $7.5^2$ deg$^2$ has been examined. The amplitude of the dipole  angular spectrum has been computed for all local $\chi^2_{\diamond-\diamond}$ (see Fig. \ref{fig:powerchi}). The probability of significance of obtaining the asymmetry amplitude, confirmed that all component separations are consistent with the symmetric map (Table \ref{tab:ass1}). The absence of asymmetry significance allowed us to avoid searching for the asymmetry direction in this paper.  However, it is useful to evaluate  the higher angular power spectrum  modes  to explore probable anomalies encoded in the clustering of local extrema, which is beyond  the scope of the current work. The \texttt{SMICA} map had different behaviour  compared to other observed maps when we applied $\Psi_{\rm tr-tr}(\vartheta\le 0.0)$, however its asymmetry was in agreement with corresponding synthetic data set.

We also derived an upper bound on the cosmic string's tension via the unweighted TPCF of peaks on the {\it Planck} data. The upper bound on the  $G\mu$ obtained by peak-peak statistics is almost higher than those bounds  derived by other methods; consequently, one concludes that the searching for cosmic strings in the CMB according to the  local extrema is very sensitive to even small residual shot-noise in the observations (Table \ref{tab:cosmic}). The minimum value of the upper bound was  for  \texttt{SMICA} map ($G\mu^{(up)} \lesssim 5.59\times 10^{-7}$).

Finally, we remark that it could be interesting to consider crossing statistics \citep{Matsubara:2003yt} and construct  a complex network from the CMB map to examine the statistical properties and explore probable exotic features such as the cosmic-string network \citep{albert2002statistical,barabasi2016network}. Topological data analysis under the  banner of the homology group is another useful approach in order to achieve  the above-mentioned purpose \citep{zomorodian2005topology11,carlsson2009topology}.  The implementation of topological and geometrical measures on polarization and convergence maps will be left for our future work.

\section*{Acknowledgements}
The authors are very grateful to Ravi K. Sheth and M. Farhang for their  extremely useful comments on different parts of this paper.  The numerical simulations were carried out on Baobab at the computing cluster of the University of Geneva. SMSM appreciates the hospitality of HECAP section of ICTP where part of this research was carried out. AVS has received funding from the European Union's Horizon 2020 research and innovation program under the Marie Sklodowska-Curie grant agreement No 674896 and No 690575 to visit Max Planck Institute for Physics in Munich. AVS also is grateful to Max Planck Institute for Physics in Munich, where part of this work was completed. 

\section*{Data Availability}
The new generated data and computational program underlying this article will be shared on reasonable request to the corresponding author.

\bibliography{pp_planck}{}

\begin{thebibliography}{}
\makeatletter
\relax
\def\mn@urlcharsother{\let\do\@makeother \do\$\do\&\do\#\do\^\do\_\do\%\do\~}
\def\mn@doi{\begingroup\mn@urlcharsother \@ifnextchar [ {\mn@doi@}
  {\mn@doi@[]}}
\def\mn@doi@[#1]#2{\def\@tempa{#1}\ifx\@tempa\@empty \href
  {http://dx.doi.org/#2} {doi:#2}\else \href {http://dx.doi.org/#2} {#1}\fi
  \endgroup}
\def\mn@eprint#1#2{\mn@eprint@#1:#2::\@nil}
\def\mn@eprint@arXiv#1{\href {http://arxiv.org/abs/#1} {{\tt arXiv:#1}}}
\def\mn@eprint@dblp#1{\href {http://dblp.uni-trier.de/rec/bibtex/#1.xml}
  {dblp:#1}}
\def\mn@eprint@#1:#2:#3:#4\@nil{\def\@tempa {#1}\def\@tempb {#2}\def\@tempc
  {#3}\ifx \@tempc \@empty \let \@tempc \@tempb \let \@tempb \@tempa \fi \ifx
  \@tempb \@empty \def\@tempb {arXiv}\fi \@ifundefined
  {mn@eprint@\@tempb}{\@tempb:\@tempc}{\expandafter \expandafter \csname
  mn@eprint@\@tempb\endcsname \expandafter{\@tempc}}}

\bibitem[\protect\citeauthoryear{Adhikari, Deutsch  \& Shandera}{Adhikari
  et~al.}{2018}]{adhikari2018statistical}
Adhikari S.,  Deutsch A.-S.,   Shandera S.,  2018, Physical Review D, 98,
  023520

\bibitem[\protect\citeauthoryear{Adler}{Adler}{1981}]{adler81}
Adler R.,  1981, The Geometry of Random Fields, Chichester: Wiley, 1981

\bibitem[\protect\citeauthoryear{Adler \& Taylor}{Adler \&
  Taylor}{2011}]{adler2011topological}
Adler R.,  Taylor J.~E.,  2011, Topological Complexity of Smooth Random
  Functions: {\'E}cole D'{\'E}t{\'e} de Probabilit{\'e}s de Saint-Flour
  XXXIX-2009.
Springer Science \& Business Media

\bibitem[\protect\citeauthoryear{Adler, Bobrowski, Borman, Subag, Weinberger
  et~al.}{Adler et~al.}{2010}]{adler2010persistent}
Adler R.~J.,  Bobrowski O.,  Borman M.~S.,  Subag E.,  Weinberger S.,   et~al.,
  2010, in , Borrowing strength: theory powering applications--a Festschrift
  for Lawrence D. Brown.
Institute of Mathematical Statistics, pp 124--143

\bibitem[\protect\citeauthoryear{Akrami, Fantaye, Shafieloo, Eriksen, Hansen,
  Banday  \& G{\'o}rski}{Akrami et~al.}{2014}]{akrami2014power}
Akrami Y.,  Fantaye Y.,  Shafieloo A.,  Eriksen H.,  Hansen F.~K.,  Banday
  A.~J.,   G{\'o}rski K.~M.,  2014, The Astrophysical Journal Letters, 784, L42

\bibitem[\protect\citeauthoryear{Albert \& Barab{\'a}si}{Albert \&
  Barab{\'a}si}{2002}]{albert2002statistical}
Albert R.,  Barab{\'a}si A.-L.,  2002, Reviews of modern physics, 74, 47

\bibitem[\protect\citeauthoryear{Bardeen, Bond, Kaiser  \& Szalay}{Bardeen
  et~al.}{1986}]{Bardeen:1985tr}
Bardeen J.~M.,  Bond J.~R.,  Kaiser N.,   Szalay A.~S.,  1986, \mn@doi
  [Astrophys. J.] {10.1086/164143}, 304, 15

\bibitem[\protect\citeauthoryear{Barreiro, Sanz, Mart{\'\i}nez-Gonz{\'a}lez,
  Cay{\'o}n  \& Silk}{Barreiro et~al.}{1997}]{Barreiro:1996ds}
Barreiro R.,  Sanz J.,  Mart{\'\i}nez-Gonz{\'a}lez E.,  Cay{\'o}n L.,   Silk
  J.,  1997, The Astrophysical Journal, 478, 1

\bibitem[\protect\citeauthoryear{Bennett \& Bouchet}{Bennett \&
  Bouchet}{1990}]{Bennett:1990}
Bennett D.~P.,  Bouchet F.~R.,  1990, Physical Review D, 41, 2408

\bibitem[\protect\citeauthoryear{Bernardeau, Colombi, Gaztanaga  \&
  Scoccimarro}{Bernardeau et~al.}{2002}]{Bernardeau:2001qr}
Bernardeau F.,  Colombi S.,  Gaztanaga E.,   Scoccimarro R.,  2002, Physics
  reports, 367, 1

\bibitem[\protect\citeauthoryear{Bevis, Hindmarsh, Kunz  \& Urrestilla}{Bevis
  et~al.}{2008}]{Bevis:2007gh}
Bevis N.,  Hindmarsh M.,  Kunz M.,   Urrestilla J.,  2008, \mn@doi [Phys. Rev.
  Lett.] {10.1103/PhysRevLett.100.021301}, 100, 021301

\bibitem[\protect\citeauthoryear{Bevis, Hindmarsh, Kunz  \& Urrestilla}{Bevis
  et~al.}{2010}]{Bevis:2010gj}
Bevis N.,  Hindmarsh M.,  Kunz M.,   Urrestilla J.,  2010, \mn@doi [Phys. Rev.]
  {10.1103/PhysRevD.82.065004}, D82, 065004

\bibitem[\protect\citeauthoryear{Bond \& Efstathiou}{Bond \&
  Efstathiou}{1987}]{Bond:1987ub}
Bond J.,  Efstathiou G.,  1987, Monthly Notices of the Royal Astronomical
  Society, 226, 655

\bibitem[\protect\citeauthoryear{Borgani}{Borgani}{1995}]{Borgani:1994uy}
Borgani S.,  1995, Physics Reports, 251, 1

\bibitem[\protect\citeauthoryear{Bouchet, Bennett  \& Stebbins}{Bouchet
  et~al.}{1988}]{Bouchet:1988hh}
Bouchet F.~R.,  Bennett D.~P.,   Stebbins A.,  1988, Nature, 335, 410

\bibitem[\protect\citeauthoryear{Brill}{Brill}{2000}]{percy00}
Brill P.~H.,  2000, CORS Bulletin, 34, 9

\bibitem[\protect\citeauthoryear{Buchert, France  \& Steiner}{Buchert
  et~al.}{2017}]{buchert2017model}
Buchert T.,  France M.~J.,   Steiner F.,  2017, Classical and Quantum Gravity,
  34, 094002

\bibitem[\protect\citeauthoryear{Carlsson}{Carlsson}{2009}]{carlsson2009topology}
Carlsson G.,  2009, Bulletin of the American Mathematical Society, 46, 255

\bibitem[\protect\citeauthoryear{Cayon \& Smoot}{Cayon \&
  Smoot}{1995}]{Cayon:1995ms}
Cayon L.,  Smoot G.,  1995, The Astrophysical Journal, 452, 487

\bibitem[\protect\citeauthoryear{Codis, Pichon, Pogosyan, Bernardeau  \&
  Matsubara}{Codis et~al.}{2013}]{Codis:2013exa}
Codis S.,  Pichon C.,  Pogosyan D.,  Bernardeau F.,   Matsubara T.,  2013,
  Monthly Notices of the Royal Astronomical Society, 435, 531

\bibitem[\protect\citeauthoryear{Cole \& Shiu}{Cole \&
  Shiu}{2018}]{cole2018persistent}
Cole A.,  Shiu G.,  2018, Journal of Cosmology and Astroparticle Physics, 2018,
  025

\bibitem[\protect\citeauthoryear{Colley \& Gott~III}{Colley \&
  Gott~III}{2015}]{Colley:2014nna}
Colley W.~N.,  Gott~III J.~R.,  2015, Monthly Notices of the Royal Astronomical
  Society, 447, 2034

\bibitem[\protect\citeauthoryear{Colley \& Richard Gott~III}{Colley \& Richard
  Gott~III}{2003}]{colley2003genus}
Colley W.~N.,  Richard Gott~III J.,  2003, Monthly Notices of the Royal
  Astronomical Society, 344, 686

\bibitem[\protect\citeauthoryear{Cooray \& Sheth}{Cooray \&
  Sheth}{2002}]{Cooray:2002dia}
Cooray A.,  Sheth R.,  2002, Physics Reports, 372, 1

\bibitem[\protect\citeauthoryear{Copeland \& Kibble}{Copeland \&
  Kibble}{2010}]{Copeland:2009ga}
Copeland E.~J.,  Kibble T. W.~B.,  2010, \mn@doi [Proc. Roy. Soc. Lond.]
  {10.1098/rspa.2009.0591}, A466, 623

\bibitem[\protect\citeauthoryear{Copeland, Liddle, Lyth, Stewart  \&
  Wands}{Copeland et~al.}{1994}]{Copeland:1994vg}
Copeland E.~J.,  Liddle A.~R.,  Lyth D.~H.,  Stewart E.~D.,   Wands D.,  1994,
  \mn@doi [Phys. Rev.] {10.1103/PhysRevD.49.6410}, D49, 6410

\bibitem[\protect\citeauthoryear{Copeland, Myers  \& Polchinski}{Copeland
  et~al.}{2004}]{Copeland:2003bj}
Copeland E.~J.,  Myers R.~C.,   Polchinski J.,  2004, \mn@doi [JHEP]
  {10.1088/1126-6708/2004/06/013}, 06, 013

\bibitem[\protect\citeauthoryear{Copi, Huterer  \& Starkman}{Copi
  et~al.}{2004}]{copi2004multipole}
Copi C.~J.,  Huterer D.,   Starkman G.~D.,  2004, Physical Review D, 70, 043515

\bibitem[\protect\citeauthoryear{Copi, Huterer, Schwarz  \& Starkman}{Copi
  et~al.}{2006}]{copi2006large}
Copi C.~J.,  Huterer D.,  Schwarz D.~J.,   Starkman G.~D.,  2006, Monthly
  Notices of the Royal Astronomical Society, 367, 79

\bibitem[\protect\citeauthoryear{Davis \& Peebles}{Davis \&
  Peebles}{1983}]{davis_peeb83}
Davis M.,  Peebles P.,  1983, The Astrophysical Journal, 267, 465

\bibitem[\protect\citeauthoryear{DePies \& Hogan}{DePies \&
  Hogan}{2007}]{Depies:2009im}
DePies M.~R.,  Hogan C.~J.,  2007, Physical Review D, 75, 125006

\bibitem[\protect\citeauthoryear{Desjacques, Crocce, Scoccimarro  \&
  Sheth}{Desjacques et~al.}{2010}]{desjacques2010modeling}
Desjacques V.,  Crocce M.,  Scoccimarro R.,   Sheth R.~K.,  2010, Physical
  Review D, 82, 103529

\bibitem[\protect\citeauthoryear{Desjacques, Jeong  \& Schmidt}{Desjacques
  et~al.}{2018}]{desjacques2018large}
Desjacques V.,  Jeong D.,   Schmidt F.,  2018, Physics reports, 733, 1

\bibitem[\protect\citeauthoryear{Dickinson}{Dickinson}{2016}]{dickinson2016cmb}
Dickinson C.,  2016, arXiv preprint arXiv:1606.03606

\bibitem[\protect\citeauthoryear{Dodelson}{Dodelson}{2003}]{dodelson2003modern}
Dodelson S.,  2003, Modern cosmology.
Academic press

\bibitem[\protect\citeauthoryear{Dor{\'e}, Colombi  \& Bouchet}{Dor{\'e}
  et~al.}{2003}]{Dore:2002xm}
Dor{\'e} O.,  Colombi S.,   Bouchet F.~R.,  2003, Monthly Notices of the Royal
  Astronomical Society, 344, 905

\bibitem[\protect\citeauthoryear{Dvali \& Vilenkin}{Dvali \&
  Vilenkin}{2004}]{Dvali:2003zj}
Dvali G.,  Vilenkin A.,  2004, \mn@doi [JCAP] {10.1088/1475-7516/2004/03/010},
  0403, 010

\bibitem[\protect\citeauthoryear{Eriksen, Hansen, Banday, G{\'o}rski  \&
  Lilje}{Eriksen et~al.}{2004}]{eriksen2004asymmetries}
Eriksen H.~K.,  Hansen F.~K.,  Banday A.~J.,  G{\'o}rski K.~M.,   Lilje P.~B.,
  2004, The Astrophysical Journal, 605, 14

\bibitem[\protect\citeauthoryear{Fabbri \& Torres}{Fabbri \&
  Torres}{1996}]{Fabbri:1995md}
Fabbri R.,  Torres S.,  1996, Astronomy and Astrophysics, 307, 703

\bibitem[\protect\citeauthoryear{Fang, Li  \& Zhao}{Fang
  et~al.}{2017}]{fang2017new}
Fang W.,  Li B.,   Zhao G.-B.,  2017, Physical review letters, 118, 181301

\bibitem[\protect\citeauthoryear{Fraisse, Ringeval, Spergel  \&
  Bouchet}{Fraisse et~al.}{2008}]{Fraisse:2007nu}
Fraisse A.~A.,  Ringeval C.,  Spergel D.~N.,   Bouchet F.~R.,  2008, Physical
  Review D, 78, 043535

\bibitem[\protect\citeauthoryear{Futamase \& Takada}{Futamase \&
  Takada}{2000}]{Futamase:2000qb}
Futamase T.,  Takada M.,  2000, arXiv preprint astro-ph/0009153

\bibitem[\protect\citeauthoryear{Gay, Pichon  \& Pogosyan}{Gay
  et~al.}{2012}]{Gay:2011wz}
Gay C.,  Pichon C.,   Pogosyan D.,  2012, Physical Review D, 85, 023011

\bibitem[\protect\citeauthoryear{Ghasemi~Nezhadhaghighi, Movahed, Yasseri  \&
  Vaez~Allaei}{Ghasemi~Nezhadhaghighi et~al.}{2017}]{sadegh15}
Ghasemi~Nezhadhaghighi M.,  Movahed S.,  Yasseri T.,   Vaez~Allaei S.~M.,
  2017, Journal of Applied Physics, 122, 085302

\bibitem[\protect\citeauthoryear{Ghasemi, Bahraminasab, Movahed, Rahvar,
  Sreenivasan  \& Tabar}{Ghasemi et~al.}{2006}]{ghasemi2006characteristic}
Ghasemi F.,  Bahraminasab A.,  Movahed M.~S.,  Rahvar S.,  Sreenivasan K.,
  Tabar M. R.~R.,  2006, Journal of Statistical Mechanics: Theory and
  Experiment, 2006, P11008

\bibitem[\protect\citeauthoryear{Gorski, Hivon, Banday, Wandelt, Hansen,
  Reinecke  \& Bartelmann}{Gorski et~al.}{2005}]{Gorski:2004by}
Gorski K.~M.,  Hivon E.,  Banday A.,  Wandelt B.~D.,  Hansen F.~K.,  Reinecke
  M.,   Bartelmann M.,  2005, The Astrophysical Journal, 622, 759

\bibitem[\protect\citeauthoryear{Gott, Colley, Park, Park  \& Mugnolo}{Gott
  et~al.}{2007}]{Gott:2006za}
Gott III J.~R.,  Colley W.~N.,  Park C.-G.,  Park C.,   Mugnolo C.,  2007,
  \mn@doi [Mon. Not. Roy. Astron. Soc.] {10.1111/j.1365-2966.2007.11730.x},
  377, 1668

\bibitem[\protect\citeauthoryear{Hadwiger}{Hadwiger}{2013}]{hadwiger2013vorlesungen}
Hadwiger H.,  2013, Vorlesungen {\"u}ber inhalt, Oberfl{\"a}che und
  isoperimetrie.
 Vol. 93, Springer-Verlag

\bibitem[\protect\citeauthoryear{Hajian \& Souradeep}{Hajian \&
  Souradeep}{2003}]{hajian2003measuring}
Hajian A.,  Souradeep T.,  2003, The Astrophysical Journal Letters, 597, L5

\bibitem[\protect\citeauthoryear{Hajian \& Souradeep}{Hajian \&
  Souradeep}{2006}]{hajian2006testing}
Hajian A.,  Souradeep T.,  2006, Physical Review D, 74, 123521

\bibitem[\protect\citeauthoryear{Hamilton}{Hamilton}{1993}]{hamilton1993toward}
Hamilton A.,  1993, The Astrophysical Journal, 417, 19

\bibitem[\protect\citeauthoryear{Hansen, Banday, G{\'o}rski, Eriksen  \&
  Lilje}{Hansen et~al.}{2009}]{hansen2009power}
Hansen F.,  Banday A.,  G{\'o}rski K.,  Eriksen H.,   Lilje P.,  2009, The
  Astrophysical Journal, 704, 1448

\bibitem[\protect\citeauthoryear{Hanson \& Lewis}{Hanson \&
  Lewis}{2009}]{hanson2009estimators}
Hanson D.,  Lewis A.,  2009, Physical Review D, 80, 063004

\bibitem[\protect\citeauthoryear{Heavens \& Gupta}{Heavens \&
  Gupta}{2001}]{Heavens:2000mu}
Heavens A.~F.,  Gupta S.,  2001, Monthly Notices of the Royal Astronomical
  Society, 324, 960

\bibitem[\protect\citeauthoryear{Heavens \& Sheth}{Heavens \&
  Sheth}{1999}]{Heavens:1999cq}
Heavens A.~F.,  Sheth R.~K.,  1999, Monthly Notices of the Royal Astronomical
  Society, 310, 1062

\bibitem[\protect\citeauthoryear{Henry~Tye}{Henry~Tye}{2008}]{HenryTye:2006uv}
Henry~Tye S.~H.,  2008, Lect. Notes Phys., 737, 949

\bibitem[\protect\citeauthoryear{Hern{\'a}ndez-Monteagudo, Kashlinsky  \&
  Atrio-Barandela}{Hern{\'a}ndez-Monteagudo
  et~al.}{2004}]{HernandezMonteagudo:2002df}
Hern{\'a}ndez-Monteagudo C.,  Kashlinsky A.,   Atrio-Barandela F.,  2004,
  Astronomy \& Astrophysics, 413, 833

\bibitem[\protect\citeauthoryear{Hewett}{Hewett}{1982}]{hewet82}
Hewett P.~C.,  1982, Monthly Notices of the Royal Astronomical Society, 201,
  867

\bibitem[\protect\citeauthoryear{Hikage, Komatsu  \& Matsubara}{Hikage
  et~al.}{2006}]{Hikage:2006fe}
Hikage C.,  Komatsu E.,   Matsubara T.,  2006, The Astrophysical Journal, 653,
  11

\bibitem[\protect\citeauthoryear{Hindmarsh \& Kibble}{Hindmarsh \&
  Kibble}{1995}]{Hindmarsh:1994re}
Hindmarsh M.~B.,  Kibble T. W.~B.,  1995, \mn@doi [Rept. Prog. Phys.]
  {10.1088/0034-4885/58/5/001}, 58, 477

\bibitem[\protect\citeauthoryear{Hoftuft, Eriksen, Banday, Gorski, Hansen  \&
  Lilje}{Hoftuft et~al.}{2009}]{hoftuft2009increasing}
Hoftuft J.,  Eriksen H.,  Banday A.,  Gorski K.,  Hansen F.,   Lilje P.,  2009,
  The Astrophysical Journal, 699, 985

\bibitem[\protect\citeauthoryear{{Hou}, {Banday}  \& {G{\'o}rski}}{{Hou}
  et~al.}{2009}]{2009MNRAS.396.1273H}
{Hou} Z.,  {Banday} A.~J.,   {G{\'o}rski} K.~M.,  2009, \mn@doi [\mnras]
  {10.1111/j.1365-2966.2009.14810.x}, \href
  {https://ui.adsabs.harvard.edu/abs/2009MNRAS.396.1273H} {396, 1273}

\bibitem[\protect\citeauthoryear{Jensen \& Szalay}{Jensen \&
  Szalay}{1986}]{jensen1986n}
Jensen L.~G.,  Szalay A.,  1986, The Astrophysical Journal, 305, L5

\bibitem[\protect\citeauthoryear{Kaiser}{Kaiser}{1984}]{kaiser1984spatial}
Kaiser N.,  1984, The Astrophysical Journal, 284, L9

\bibitem[\protect\citeauthoryear{Kashlinsky}{Kashlinsky}{2005}]{Kashlinsky:2004jt}
Kashlinsky A.,  2005, Physics Reports, 409, 361

\bibitem[\protect\citeauthoryear{Kashlinsky, Hern{\'a}ndez-Monteagudo  \&
  Atrio-Barandela}{Kashlinsky et~al.}{2001}]{Kashlinsky:2001zla}
Kashlinsky A.,  Hern{\'a}ndez-Monteagudo C.,   Atrio-Barandela F.,  2001, The
  Astrophysical Journal Letters, 557, L1

\bibitem[\protect\citeauthoryear{Kerscher, Szapudi  \& Szalay}{Kerscher
  et~al.}{2000}]{kerscher2000comparison}
Kerscher M.,  Szapudi I.,   Szalay A.~S.,  2000, The Astrophysical Journal
  Letters, 535, L13

\bibitem[\protect\citeauthoryear{Kibble}{Kibble}{1976}]{Kibble:1976sj}
Kibble T. W.~B.,  1976, \mn@doi [J. Phys.] {10.1088/0305-4470/9/8/029}, A9,
  1387

\bibitem[\protect\citeauthoryear{Kibble}{Kibble}{1980}]{Kibble:1980mv22}
Kibble T. W.~B.,  1980, \mn@doi [Phys. Rept.] {10.1016/0370-1573(80)90091-5},
  67, 183

\bibitem[\protect\citeauthoryear{Kibble}{Kibble}{2004}]{Kibble:2004hq}
Kibble T.~W.,  2004, arXiv preprint astro-ph/0410073

\bibitem[\protect\citeauthoryear{Kogut, Banday, Bennett, Hinshaw, Lubin  \&
  Smoot}{Kogut et~al.}{1995}]{kogut95}
Kogut A.~J.,  Banday A.~J.,  Bennett C.~L.,  Hinshaw G.~F.,  Lubin P.~M.,
  Smoot G.~F.,  1995, Astrophysical Journal, 439, 29

\bibitem[\protect\citeauthoryear{Kogut, Banday, Bennett, G{\'o}rski, Hinshaw,
  Smoot  \& Wright}{Kogut et~al.}{1996}]{kogut96}
Kogut A.,  Banday A.~J.,  Bennett C.~L.,  G{\'o}rski K.~M.,  Hinshaw G.,  Smoot
  G.~F.,   Wright E.~L.,  1996, The Astrophysical Journal Letters, 464, L29

\bibitem[\protect\citeauthoryear{Komatsu}{Komatsu}{2002}]{komatsu2002pursuit}
Komatsu E.,  2002, arXiv preprint astro-ph/0206039

\bibitem[\protect\citeauthoryear{Landy \& Szalay}{Landy \&
  Szalay}{1993}]{Landy:1993yu}
Landy S.~D.,  Szalay A.~S.,  1993, \mn@doi [Astrophys. J.] {10.1086/172900},
  412, 64

\bibitem[\protect\citeauthoryear{{Larson} \& {Wandelt}}{{Larson} \&
  {Wandelt}}{2004}]{larson2004hot}
{Larson} D.~L.,  {Wandelt} B.~D.,  2004, \mn@doi [\apjl] {10.1086/425250},
  \href {https://ui.adsabs.harvard.edu/abs/2004ApJ...613L..85L} {613, L85}

\bibitem[\protect\citeauthoryear{Larson \& Wandelt}{Larson \&
  Wandelt}{2005}]{Larson:2005vb}
Larson D.~L.,  Wandelt B.~D.,  2005, arXiv preprint astro-ph/0505046

\bibitem[\protect\citeauthoryear{Lesgourges}{Lesgourges}{2013}]{Lesgourgues:2013qba}
Lesgourges J.,  2013, in , Searching for New Physics at Small and Large Scales:
  TASI 2012.
World Scientific, pp 29--97

\bibitem[\protect\citeauthoryear{Lesgourgues, Mangano, Miele  \&
  Pastor}{Lesgourgues et~al.}{2013}]{lesg13}
Lesgourgues J.,  Mangano G.,  Miele G.,   Pastor S.,  2013, Neutrino cosmology.
Cambridge University Press

\bibitem[\protect\citeauthoryear{Lewis \& Challinor}{Lewis \&
  Challinor}{2006}]{Lewis:2006fu}
Lewis A.,  Challinor A.,  2006, Physics Reports, 429, 1

\bibitem[\protect\citeauthoryear{Ling, Wang, Li, Li, Wang  \& Gao}{Ling
  et~al.}{2015}]{ling2015distinguishing}
Ling C.,  Wang Q.,  Li R.,  Li B.,  Wang J.,   Gao L.,  2015, Physical Review
  D, 92, 064024

\bibitem[\protect\citeauthoryear{Lumsden, Heavens  \& Peacock}{Lumsden
  et~al.}{1989}]{lumusden89}
Lumsden S.,  Heavens A.,   Peacock J.,  1989, Monthly Notices of the Royal
  Astronomical Society, 238, 293

\bibitem[\protect\citeauthoryear{Majumdar \& Christine-Davis}{Majumdar \&
  Christine-Davis}{2002}]{Majumdar:2002hy}
Majumdar M.,  Christine-Davis A.,  2002, \mn@doi [JHEP]
  {10.1088/1126-6708/2002/03/056}, 03, 056

\bibitem[\protect\citeauthoryear{Malik \& Wands}{Malik \&
  Wands}{2009}]{Malik:2008im}
Malik K.~A.,  Wands D.,  2009, \mn@doi [Phys. Rept.]
  {10.1016/j.physrep.2009.03.001}, 475, 1

\bibitem[\protect\citeauthoryear{Marcos-Caballero, Fern\'andez-Cobos,
  Mart\'\i{}nez-Gonz\'alez  \& Vielva}{Marcos-Caballero
  et~al.}{2016}]{Marcos-Caballero:2015lxp}
Marcos-Caballero A.,  Fern\'andez-Cobos R.,  Mart\'\i{}nez-Gonz\'alez E.,
  Vielva P.,  2016, \mn@doi [JCAP] {10.1088/1475-7516/2016/04/058}, 04, 058

\bibitem[\protect\citeauthoryear{Marques, Novaes, Bernui  \& Ferreira}{Marques
  et~al.}{2018}]{marques2018isotropy}
Marques G.,  Novaes C.,  Bernui A.,   Ferreira I.,  2018, Monthly Notices of
  the Royal Astronomical Society, 473, 165

\bibitem[\protect\citeauthoryear{Martinez \& Saar}{Martinez \&
  Saar}{2001}]{martinez2001statistics}
Martinez V.~J.,  Saar E.,  2001, Statistics of the galaxy distribution.
CRC press

\bibitem[\protect\citeauthoryear{Matsubara}{Matsubara}{1996}]{mat96a}
Matsubara T.,  1996, The Astrophysical Journal, 457, 13

\bibitem[\protect\citeauthoryear{Matsubara}{Matsubara}{2003}]{Matsubara:2003yt}
Matsubara T.,  2003, The Astrophysical Journal, 584, 1

\bibitem[\protect\citeauthoryear{Matsubara}{Matsubara}{2010}]{matsubara2010analytic}
Matsubara T.,  2010, Physical Review D, 81, 083505

\bibitem[\protect\citeauthoryear{Matsubara}{Matsubara}{2020}]{matsubara2020statistics}
Matsubara T.,  2020, Physical Review D, 101, 043532

\bibitem[\protect\citeauthoryear{Mecke, Bucheri  \& Wagner}{Mecke
  et~al.}{1994}]{mecke1994robust}
Mecke K.,  Bucheri T.,   Wagner H.,  1994, Astron. Astrophys, 288, 697

\bibitem[\protect\citeauthoryear{Movahed \& Khosravi}{Movahed \&
  Khosravi}{2011}]{sadegh11}
Movahed M.~S.,  Khosravi S.,  2011, Journal of Cosmology and Astroparticle
  Physics, 2011, 012

\bibitem[\protect\citeauthoryear{Movahed, Ghasemi, Rahvar  \& Tabar}{Movahed
  et~al.}{2011}]{SadeghMovahed:2006em}
Movahed M.~S.,  Ghasemi F.,  Rahvar S.,   Tabar M. R.~R.,  2011, Physical
  Review E, 84, 021103

\bibitem[\protect\citeauthoryear{Movahed, Javanmardi  \& Sheth}{Movahed
  et~al.}{2013}]{Movahed:2012zt}
Movahed M.~S.,  Javanmardi B.,   Sheth R.~K.,  2013, Monthly Notices of the
  Royal Astronomical Society, 434, 3597

\bibitem[\protect\citeauthoryear{Novaes, Bernui, Marques  \& Ferreira}{Novaes
  et~al.}{2016}]{novaes2016local}
Novaes C.,  Bernui A.,  Marques G.,   Ferreira I.,  2016, Monthly Notices of
  the Royal Astronomical Society, 461, 1363

\bibitem[\protect\citeauthoryear{Peacock \& Heavens}{Peacock \&
  Heavens}{1985}]{peac85}
Peacock J.,  Heavens A.~F.,  1985, Monthly Notices of the Royal Astronomical
  Society, 217, 805

\bibitem[\protect\citeauthoryear{Peebles}{Peebles}{1980}]{peeb80}
Peebles P. J.~E.,  1980, The large-scale structure of the universe.
Princeton university press

\bibitem[\protect\citeauthoryear{{Planck Collaboration} et~al.,}{{Planck
  Collaboration} et~al.}{2014a}]{planck2013results}
{Planck Collaboration} et~al., 2014a, \mn@doi [\aap]
  {10.1051/0004-6361/201321534}, \href
  {https://ui.adsabs.harvard.edu/abs/2014A&A...571A..23P} {571, A23}

\bibitem[\protect\citeauthoryear{{Planck Collaboration} et~al.,}{{Planck
  Collaboration} et~al.}{2014b}]{ade2014plancknon}
{Planck Collaboration} et~al., 2014b, \mn@doi [\aap]
  {10.1051/0004-6361/201321554}, \href
  {https://ui.adsabs.harvard.edu/abs/2014A&A...571A..24P} {571, A24}

\bibitem[\protect\citeauthoryear{{Planck Collaboration} et~al.,}{{Planck
  Collaboration} et~al.}{2014c}]{Ade:2013xla}
{Planck Collaboration} et~al., 2014c, \mn@doi [\aap]
  {10.1051/0004-6361/201321621}, \href
  {https://ui.adsabs.harvard.edu/abs/2014A&A...571A..25P} {571, A25}

\bibitem[\protect\citeauthoryear{{Planck Collaboration} et~al.,}{{Planck
  Collaboration} et~al.}{2016a}]{2016A&A...594A...9P}
{Planck Collaboration} et~al., 2016a, \mn@doi [\aap]
  {10.1051/0004-6361/201525936}, \href
  {https://ui.adsabs.harvard.edu/abs/2016A&A...594A...9P} {594, A9}

\bibitem[\protect\citeauthoryear{{Planck Collaboration} et~al.,}{{Planck
  Collaboration} et~al.}{2016b}]{Ade:2015xua}
{Planck Collaboration} et~al., 2016b, \mn@doi [\aap]
  {10.1051/0004-6361/201525830}, \href
  {https://ui.adsabs.harvard.edu/abs/2016A&A...594A..13P} {594, A13}

\bibitem[\protect\citeauthoryear{{Planck Collaboration} et~al.,}{{Planck
  Collaboration} et~al.}{2016c}]{Ade:2015hxq}
{Planck Collaboration} et~al., 2016c, \mn@doi [\aap]
  {10.1051/0004-6361/201526681}, \href
  {https://ui.adsabs.harvard.edu/abs/2016A&A...594A..16P} {594, A16}

\bibitem[\protect\citeauthoryear{{Planck Collaboration} et~al.,}{{Planck
  Collaboration} et~al.}{2016d}]{2016A&A...594A..17P}
{Planck Collaboration} et~al., 2016d, \mn@doi [\aap]
  {10.1051/0004-6361/201525836}, \href
  {https://ui.adsabs.harvard.edu/abs/2016A&A...594A..17P} {594, A17}

\bibitem[\protect\citeauthoryear{{Planck Collaboration} et~al.,}{{Planck
  Collaboration} et~al.}{2020a}]{plancke2elow}
{Planck Collaboration} et~al., 2020a, \mn@doi [\aap]
  {10.1051/0004-6361/201833293}, \href
  {https://ui.adsabs.harvard.edu/abs/2020A&A...641A...2P} {641, A2}

\bibitem[\protect\citeauthoryear{{Planck Collaboration} et~al.,}{{Planck
  Collaboration} et~al.}{2020b}]{Aghanim:2018fcm}
{Planck Collaboration} et~al., 2020b, \mn@doi [\aap]
  {10.1051/0004-6361/201832909}, \href
  {https://ui.adsabs.harvard.edu/abs/2020A&A...641A...3P} {641, A3}

\bibitem[\protect\citeauthoryear{{Planck Collaboration} et~al.,}{{Planck
  Collaboration} et~al.}{2020c}]{akrami2020planckcomponent}
{Planck Collaboration} et~al., 2020c, \mn@doi [\aap]
  {10.1051/0004-6361/201833881}, \href
  {https://ui.adsabs.harvard.edu/abs/2020A&A...641A...4P} {641, A4}

\bibitem[\protect\citeauthoryear{{Planck Collaboration} et~al.,}{{Planck
  Collaboration} et~al.}{2020d}]{Aghanim:2018eyx}
{Planck Collaboration} et~al., 2020d, \mn@doi [\aap]
  {10.1051/0004-6361/201833910}, \href
  {https://ui.adsabs.harvard.edu/abs/2020A&A...641A...6P} {641, A6}

\bibitem[\protect\citeauthoryear{{Planck Collaboration} et~al.,}{{Planck
  Collaboration} et~al.}{2020e}]{akrami2020planck}
{Planck Collaboration} et~al., 2020e, \mn@doi [\aap]
  {10.1051/0004-6361/201935201}, \href
  {https://ui.adsabs.harvard.edu/abs/2020A&A...641A...7P} {641, A7}

\bibitem[\protect\citeauthoryear{{Planck Collaboration} et~al.,}{{Planck
  Collaboration} et~al.}{2020f}]{akrami2020plancknong}
{Planck Collaboration} et~al., 2020f, \mn@doi [\aap]
  {10.1051/0004-6361/201935891}, \href
  {https://ui.adsabs.harvard.edu/abs/2020A&A...641A...9P} {641, A9}

\bibitem[\protect\citeauthoryear{Pogosian, Tye, Wasserman  \& Wyman}{Pogosian
  et~al.}{2003}]{Pogosian:2003mz}
Pogosian L.,  Tye S.-H.~H.,  Wasserman I.,   Wyman M.,  2003, Physical Review
  D, 68, 023506

\bibitem[\protect\citeauthoryear{Pogosyan, Pichon, Gay, Prunet, Cardoso,
  Sousbie  \& Colombi}{Pogosyan et~al.}{2009}]{Pogosyan:2008jb}
Pogosyan D.,  Pichon C.,  Gay C.,  Prunet S.,  Cardoso J.,  Sousbie T.,
  Colombi S.,  2009, Monthly Notices of the Royal Astronomical Society, 396,
  635

\bibitem[\protect\citeauthoryear{Pogosyan, Pichon  \& Gay}{Pogosyan
  et~al.}{2011}]{Pogosyan:2011qq}
Pogosyan D.,  Pichon C.,   Gay C.,  2011, Physical Review D, 84, 083510

\bibitem[\protect\citeauthoryear{Polchinski}{Polchinski}{2005}]{Polchinski:2004hb}
Polchinski J.,  2005, \mn@doi [Int. J. Mod. Phys.] {10.1142/S0217751X05026686},
  A20, 3413

\bibitem[\protect\citeauthoryear{Politzer \& Wise}{Politzer \&
  Wise}{1984}]{politzer1984relations}
Politzer H.~D.,  Wise M.~B.,  1984, The Astrophysical Journal, 285, L1

\bibitem[\protect\citeauthoryear{Pranav et~al.,}{Pranav
  et~al.}{2019}]{pranav2019topology}
Pranav P.,  et~al., 2019, Monthly Notices of the Royal Astronomical Society,
  485, 4167

\bibitem[\protect\citeauthoryear{Prunet, Uzan, Bernardeau  \& Brunier}{Prunet
  et~al.}{2005}]{prunet2005constraints}
Prunet S.,  Uzan J.-P.,  Bernardeau F.,   Brunier T.,  2005, Physical Review D,
  71, 083508

\bibitem[\protect\citeauthoryear{Rath \& Jain}{Rath \&
  Jain}{2013}]{rath2013testing}
Rath P.~K.,  Jain P.,  2013, Journal of Cosmology and Astroparticle Physics,
  2013, 014

\bibitem[\protect\citeauthoryear{Rath, Mudholkar, Jain, Aluri  \& Panda}{Rath
  et~al.}{2013}]{rath2013direction}
Rath P.~K.,  Mudholkar T.,  Jain P.,  Aluri P.~K.,   Panda S.,  2013, Journal
  of Cosmology and Astroparticle Physics, 2013, 007

\bibitem[\protect\citeauthoryear{Rath, Samal, Panda, Mishra  \& Aluri}{Rath
  et~al.}{2018}]{rath2018testing}
Rath P.~K.,  Samal P.~K.,  Panda S.,  Mishra D.~D.,   Aluri P.~K.,  2018,
  Monthly Notices of the Royal Astronomical Society, 475, 4357

\bibitem[\protect\citeauthoryear{Reischke, Maturi  \& Bartelmann}{Reischke
  et~al.}{2015}]{Reischke:2015jga}
Reischke R.,  Maturi M.,   Bartelmann M.,  2015, Monthly Notices of the Royal
  Astronomical Society, 456, 641

\bibitem[\protect\citeauthoryear{Renaux-Petel}{Renaux-Petel}{2015}]{Renaux-Petel:2015bja}
Renaux-Petel S.,  2015, Comptes Rendus Physique, 16, 969

\bibitem[\protect\citeauthoryear{Rice}{Rice}{1944}]{rice44a}
Rice S.~O.,  1944, Bell Labs Technical Journal, 23, 282

\bibitem[\protect\citeauthoryear{Rice}{Rice}{1945}]{rice44b}
Rice S.~O.,  1945, The Bell System Technical Journal, 24, 46

\bibitem[\protect\citeauthoryear{Rice}{Rice}{1954}]{rice1954selected}
Rice S.,  1954, ed. N. Wax, Dover Publ. Inc.(NY)

\bibitem[\protect\citeauthoryear{Ringeval \& Bouchet}{Ringeval \&
  Bouchet}{2012}]{Ringeval:2012tk}
Ringeval C.,  Bouchet F.~R.,  2012, Physical Review D, 86, 023513

\bibitem[\protect\citeauthoryear{Ringeval, Sakellariadou  \& Bouchet}{Ringeval
  et~al.}{2007}]{Ringeval:2005kr}
Ringeval C.,  Sakellariadou M.,   Bouchet F.~R.,  2007, Journal of Cosmology
  and Astroparticle Physics, 2007, 023

\bibitem[\protect\citeauthoryear{Rossi}{Rossi}{2013}]{Rossi:2013fea}
Rossi G.,  2013, Monthly Notices of the Royal Astronomical Society, 430, 1486

\bibitem[\protect\citeauthoryear{Rossi, Sheth, Park  \&
  Hern{\'a}ndez-Monteagudo}{Rossi et~al.}{2009}]{Rossi:2009wm}
Rossi G.,  Sheth R.~K.,  Park C.,   Hern{\'a}ndez-Monteagudo C.,  2009, Monthly
  Notices of the Royal Astronomical Society, 399, 304

\bibitem[\protect\citeauthoryear{Rossi, Chingangbam  \& Park}{Rossi
  et~al.}{2011}]{Rossi:2010hu}
Rossi G.,  Chingangbam P.,   Park C.,  2011, Monthly Notices of the Royal
  Astronomical Society, 411, 1880

\bibitem[\protect\citeauthoryear{Ryden}{Ryden}{1988}]{ryden1988}
Ryden B.~S.,  1988, The Astrophysical Journal, 333, L41

\bibitem[\protect\citeauthoryear{Ryden, Melott, Craig, Gott~III, Weinberg,
  Scherrer, Bhavsar  \& Miller}{Ryden et~al.}{1989}]{ryd89}
Ryden B.~S.,  Melott A.~L.,  Craig D.~A.,  Gott~III J.~R.,  Weinberg D.~H.,
  Scherrer R.~J.,  Bhavsar S.~P.,   Miller J.~M.,  1989, The Astrophysical
  Journal, 340, 647

\bibitem[\protect\citeauthoryear{Sakellariadou}{Sakellariadou}{1997}]{Sakellariadou:1997zt}
Sakellariadou M.,  1997, \mn@doi [Int. J. Theor. Phys.] {10.1007/BF02768939},
  36, 2503

\bibitem[\protect\citeauthoryear{Sakellariadou}{Sakellariadou}{2007}]{Sakellariadou:2006qs}
Sakellariadou M.,  2007, \mn@doi [Lect. Notes Phys.]
  {10.1007/3-540-70859-6_10}, 718, 247

\bibitem[\protect\citeauthoryear{Sarangi \& Tye}{Sarangi \&
  Tye}{2002}]{Sarangi:2002yt}
Sarangi S.,  Tye S. H.~H.,  2002, \mn@doi [Phys. Lett.]
  {10.1016/S0370-2693(02)01824-5}, B536, 185

\bibitem[\protect\citeauthoryear{Sazhin}{Sazhin}{1985}]{sazhin1985hot}
Sazhin M.,  1985, Monthly Notices of the Royal Astronomical Society, 216, 25P

\bibitem[\protect\citeauthoryear{Schmalzing \& G{\'o}rski}{Schmalzing \&
  G{\'o}rski}{1998}]{schmalzing1998minkowski}
Schmalzing J.,  G{\'o}rski K.~M.,  1998, Monthly Notices of the Royal
  Astronomical Society, 297, 355

\bibitem[\protect\citeauthoryear{Schmalzing, Buchert  \& Kerscher}{Schmalzing
  et~al.}{1995}]{schmalzing1995minkowski}
Schmalzing J.,  Buchert T.,   Kerscher M.,  1995, Proc. Int. Sch. Phys. Fermi,
  132, 281

\bibitem[\protect\citeauthoryear{Shahbazi, Sobhanian, Tabar, Khorram, Frootan
  \& Zahed}{Shahbazi et~al.}{2003}]{tabar03}
Shahbazi F.,  Sobhanian S.,  Tabar M. R.~R.,  Khorram S.,  Frootan G.,   Zahed
  H.,  2003, Journal of Physics A: Mathematical and General, 36, 2517

\bibitem[\protect\citeauthoryear{Shellard}{Shellard}{1987}]{Shellard:1987bv}
Shellard E. P.~S.,  1987, \mn@doi [Nucl. Phys.] {10.1016/0550-3213(87)90290-2},
  B283, 624

\bibitem[\protect\citeauthoryear{Szalay}{Szalay}{1988a}]{szalay88}
Szalay A.~S.,  1988a, in Symposium-International Astronomical Union. pp
  163--167

\bibitem[\protect\citeauthoryear{Szalay}{Szalay}{1988b}]{szalay1988constraints1}
Szalay A.~S.,  1988b, The Astrophysical Journal, 333, 21

\bibitem[\protect\citeauthoryear{Takada \& Futamase}{Takada \&
  Futamase}{2001}]{takada2001detectability}
Takada M.,  Futamase T.,  2001, The Astrophysical Journal, 546, 620

\bibitem[\protect\citeauthoryear{Takada, Komatsu  \& Futamase}{Takada
  et~al.}{2000}]{takada2000gravitational}
Takada M.,  Komatsu E.,   Futamase T.,  2000, The Astrophysical Journal
  Letters, 533, L83

\bibitem[\protect\citeauthoryear{Taqqu}{Taqqu}{1977}]{taqqu1977law}
Taqqu M.~S.,  1977, Probability Theory and Related Fields, 40, 203

\bibitem[\protect\citeauthoryear{Tojeiro, Castro, Heavens  \& Gupta}{Tojeiro
  et~al.}{2006}]{Tojeiro:2005mt}
Tojeiro R.,  Castro P.,  Heavens A.,   Gupta S.,  2006, Monthly Notices of the
  Royal Astronomical Society, 365, 265

\bibitem[\protect\citeauthoryear{Vachaspati \& Vilenkin}{Vachaspati \&
  Vilenkin}{1984}]{Vachaspati:1984dz}
Vachaspati T.,  Vilenkin A.,  1984, Physical Review D, 30, 2036

\bibitem[\protect\citeauthoryear{Vafaei~Sadr, Movahed, Farhang, Ringeval  \&
  Bouchet}{Vafaei~Sadr et~al.}{2017}]{vafaei2017multiscale}
Vafaei~Sadr A.,  Movahed S.,  Farhang M.,  Ringeval C.,   Bouchet F.,  2017,
  Monthly Notices of the Royal Astronomical Society, 475, 1010

\bibitem[\protect\citeauthoryear{Vafaei~Sadr, Farhang, Movahed, Bassett  \&
  Kunz}{Vafaei~Sadr et~al.}{2018}]{sadr2018cosmic}
Vafaei~Sadr A.,  Farhang M.,  Movahed S.,  Bassett B.,   Kunz M.,  2018,
  Monthly Notices of the Royal Astronomical Society, 478, 1132

\bibitem[\protect\citeauthoryear{Vilenkin}{Vilenkin}{1981}]{Vilenkin:1981iu}
Vilenkin A.,  1981, \mn@doi [Phys. Rev. Lett.] {10.1103/PhysRevLett.46.1169,
  10.1103/PhysRevLett.46.1496}, 46, 1169

\bibitem[\protect\citeauthoryear{Vilenkin}{Vilenkin}{1985}]{Vilenkin:1984ib}
Vilenkin A.,  1985, \mn@doi [Phys. Rept.] {10.1016/0370-1573(85)90033-X}, 121,
  263

\bibitem[\protect\citeauthoryear{Vilenkin \& Shellard}{Vilenkin \&
  Shellard}{2000}]{Vilenkin:2000jqa}
Vilenkin A.,  Shellard E. P.~S.,  2000, Cosmic strings and other topological
  defects.
Cambridge University Press

\bibitem[\protect\citeauthoryear{Vittorio \& Juszkiewicz}{Vittorio \&
  Juszkiewicz}{1987}]{nicola87}
Vittorio N.,  Juszkiewicz R.,  1987, The Astrophysical Journal, 314, L29

\bibitem[\protect\citeauthoryear{Zeldovich}{Zeldovich}{1980}]{Zeldovich:1980gh}
Zeldovich {\relax Ya}.~B.,  1980, Mon. Not. Roy. Astron. Soc., 192, 663

\bibitem[\protect\citeauthoryear{Zomorodian}{Zomorodian}{2005}]{zomorodian2005topology11}
Zomorodian A.~J.,  2005, Topology for computing.
Cambridge university press

\bibitem[\protect\citeauthoryear{Zomorodian}{Zomorodian}{2016}]{barabasi2016network}
Zomorodian A.~J.,  2016, Network science.
Cambridge university press

\makeatother
\end{thebibliography}
\bibliographystyle{mnras}

\appendix
\section{Complementary definitions}\label{app1}

In this appendix, we give some complementary definitions used for computing critical sets of the CMB field.  The covariance matrix which is represented by Eq. (\ref{parti1}) is given by:
\begin{widetext}
\begin{eqnarray}\label{cov1}
\mathcal{K}^{(2)}\equiv \langle {\mathcal A} \otimes{\mathcal A}\rangle =\left[\begin{array}{cccccc}
\langle \delta_T^2\rangle & \langle \delta_T\eta_{\phi}\rangle & \langle \delta_T\eta_{\theta}\rangle &  \langle \delta_T\xi_{\phi\phi}\rangle &  \langle \delta_T\xi_{\theta\theta}\rangle&  \langle \delta_T\xi_{\phi\theta}\rangle\\
 \langle \delta_T\eta_{\phi}\rangle & \langle \eta_{\phi}^2\rangle & \langle \eta{\phi}\eta_{\theta}\rangle &  \langle \eta{\phi}\xi_{\phi\phi}\rangle &  \langle\eta{\phi}\xi_{\theta\theta}\rangle&  \langle \eta{\phi}\xi_{\phi\theta}\rangle\\
\langle \delta_T\eta_{\theta}\rangle & \langle\eta_{\theta} \eta_{\phi}\rangle & \langle\eta_{\theta}^2\rangle &  \langle \eta_{\theta}\xi_{\phi\phi}\rangle &  \langle\eta_{\theta}\xi_{\theta\theta}\rangle&  \langle \eta_{\theta}\xi_{\phi\theta}\rangle\\
\langle \delta_T\xi_{\phi\phi}\rangle & \langle\xi_{\phi\phi} \eta_{\phi}\rangle & \langle\xi_{\phi\phi}\eta_{\theta}\rangle &  \langle \xi_{\phi\phi}^2\rangle &  \langle\xi_{\phi\phi}\xi_{\theta\theta}\rangle&  \langle \xi_{\phi\phi}\xi_{\phi\theta}\rangle\\
\langle \delta_T\xi_{\theta\theta}\rangle & \langle\xi_{\theta\theta} \eta_{\phi}\rangle & \langle\xi_{\theta\theta}\eta_{\theta}\rangle &  \langle \xi_{\theta\theta}\xi_{\phi\phi}\rangle &  \langle\xi_{\theta\theta}^2\rangle&  \langle \xi_{\theta\theta}\xi_{\phi\theta}\rangle\\
\langle \delta_T\xi_{\phi\theta}\rangle & \langle\xi_{\phi\theta} \eta_{\phi}\rangle & \langle\xi_{\phi\theta}\eta_{\theta}\rangle &  \langle \xi_{\phi\theta}\xi_{\phi\phi}\rangle &  \langle\xi_{\phi\theta}\xi_{\theta\theta}\rangle&  \langle\xi_{\phi\theta}^2\rangle\\
  \end{array} \right]
\end{eqnarray}
\end{widetext}
 The non-zero elements of $\mathcal{K}^{(2)}$ for separation angle, $\cos(\psi)=|\hat{n}_i.\hat{n}_j|=1$ are as follows \citep{Bond:1987ub}:
\begin{eqnarray}
\sigma_0^2&\equiv& \langle \delta_T^2\rangle=\sum_{\ell}\frac{(2\ell+1)}{4\pi}C^{TT}_{\ell}W^2_{\ell}\\\nonumber
\langle \delta_T\xi_{\phi\phi}\rangle&=& \langle \delta_T\xi_{\theta\theta}\rangle=-\sum_{\ell} \frac{(2\ell+1)}{4\pi}\frac{\ell(\ell+1)}{2}C^{TT}_{\ell}W^2_{\ell}\\\nonumber
\langle\xi_{\theta\theta}^2\rangle&=&\sum_{\ell} \frac{(2\ell+1)}{4\pi}\frac{(3\ell(\ell+1)-2)\ell(\ell+1)}{8}C^{TT}_{\ell}W^2_{\ell}\\\nonumber
\langle \xi_{\theta\theta}\xi_{\phi\phi}\rangle &=& \sum_{\ell} \frac{(2\ell+1)}{4\pi}\frac{(\ell(\ell+1)+2)\ell(\ell+1)}{8}C^{TT}_{\ell}W^2_{\ell}\\\nonumber
 \langle\xi_{\phi\theta}^2\rangle&=& \sum_{\ell} \frac{(2\ell+1)}{4\pi}\frac{(\ell+2)(\ell+1)\ell(\ell-1)}{8}C^{TT}_{\ell}W^2_{\ell}
\end{eqnarray}
where $W_{\ell}=\exp\left(-\theta_{\rm beam}^2\ell(\ell+1)/2\right)$
and $\theta_{\rm beam}\equiv\theta_{\rm FWHM}/\sqrt{8\ln (2)}$  for
a Gaussian smoothing kernel associated with beam transfer function
\citep{Bond:1987ub,Heavens:1999cq,Hikage:2006fe}. Also
$C^{TT}_{\ell}$ is the power spectrum of CMB temperature
fluctuations. Other terms are $\langle \eta_{\phi}^2\rangle =
\langle\eta_{\theta}^2\rangle=-\langle
\delta_T\xi_{\phi\phi}\rangle$ and
$\langle\xi_{\theta\theta}^2\rangle=\langle\xi_{\phi\phi}^2\rangle$.
\\
\\
\bsp    
\label{lastpage}
\end{document}